\documentclass[utf8]{frontiersFPHY} 

\setcitestyle{square} 
\usepackage{url,hyperref,lineno,microtype,subcaption}
\usepackage{lscape}
\usepackage[onehalfspacing]{setspace}
\usepackage{caption}
\usepackage{subcaption}
\usepackage[figuresright]{rotating}
\usepackage{adjustbox}

\def\keyFont{\fontsize{8}{11}\helveticabold }
\def\firstAuthorLast{Ragusa {et~al.}} 

\def\Authors{Rossella Ragusa\,$^{1,2,*}$, Marco Mirabile\,$^{2,3}$, Marilena Spavone\,$^{1}$, Michele Cantiello\,$^{3}$, Enrichetta Iodice\,$^{1}$, Antonio La Marca\,$^{4,5}$, Maurizio Paolillo\,$^{1,2}$ and Pietro Schipani\,$^{1}$}
\def\Address{$^{1}$ INAF - Astronomical Observatory of Capodimonte,
 Salita Moiariello 16, 80131, Naples, Italy \\
$^{2}$ University of Naples Federico II, C.U. Monte Sant'Angelo, Via Cinthia, 80126, Naples, Italy \\
$^{3}$ INAF - Astronomical Observatory of Abruzzo, Via Maggini, 64100, Teramo, Italy \\
$^{4}$ SRON Netherlands Institute for Space Research, Landleven 12, 9747 AD Groningen, The Netherlands\\
$^{5}$Kapteyn Institute, University of Groningen, Landleven 12, 9747 AD Groningen, the Netherlands}

\def\corrAuthor{Rossella Ragusa}

\def\corrEmail{rossella.ragusa@inaf.it}

\begin{document}
\onecolumn
\firstpage{1}

\title[The IGL in the LEO I pair]{The intra-group baryons in the LEO I pair from the VST Early-type GAlaxy Survey} 

\author[\firstAuthorLast ]{\Authors} 
\address{} 
\correspondance{} 

\extraAuth{}

\maketitle

\begin{abstract}
\tiny
\section{In this paper we present the deep, wide-field and multi-band imaging of the LEO I pair NGC~3379-NGC~3384, from the  VST Early-type GAlaxy Survey (VEGAS).
The main goal of this study is to map the intra-group baryons in the pair, in the form of  
diffuse light and globular clusters (GCs).
Taking advantage from the large covered area, which extends for $\sim$ 3.9 square degrees around the 
pair, and the long integration time, we can map the light distribution out to $\sim$ 63 kpc and 
down to $\sim$ 30 mag/arcsec$^2$ in the $g$ band and $\sim$ 29 mag/arcsec$^2$ in the $r$ band, 
deeper than previous data available for this target. 
The map of the intra-group light (IGL) presents two very faint ($\mu_g \sim$ 28-29 mag/arcsec$^2$) streams 
protruding from the brightest group member NGC~3379 and elongated toward North-West and South. 
We estimate that the fraction of the stellar halo around NGC~3379 plus the  IGL is 
$\sim 17 \pm 2\%$  in both $g$ and $r$ bands, with an average color  $g$-$r$= $0.75 \pm 0.04$~mag. 
The color distribution of the GCs appears multi-modal, with two dominant peaks at (u-r) = 1.8 mag
and (u-r) = 2.1 mag, respectively. 
The GC population stretches from North-East to South-West and from North-West to South of the pair, in the last case overlapping with the streams of IGL,
as well as the PNe distribution found by \citet{Hartke2020} and \citet{Hartke2022arXiv220108710H}. 
Since these structures are elongated in the direction of the two nearby galaxies M96 and 
NGC~3338, they could be the remnant of a past gravitational interactions with the pair.
}

\tiny
\keyFont{\section{Keywords:} Galaxies: evolution - Galaxies: surface photometry - Galaxies:  group: general - Galaxies: interactions- intergalactic medium - Galaxies: diffuse intragroup light}

\end{abstract}

\section{Introduction}\label{sec:intro}

It has been known for many years that in the $\Lambda$CDM model the cosmic structures are formed and evolve according to the {\it hierarchical accretion} scenario. With this paradigm, the latest and largest gravitationally bound systems in the universe, i.e. galaxy clusters, are the result of past merging events of smaller elements, the galaxy groups \citep{deLucia2006}, that differ from the galaxy clusters, in the local universe, for the different values of viral masses ($M_{vir}$), which range from $\sim$ 10\textsuperscript{13} $M_{\odot}$ 
for galaxy groups to $\sim$  10\textsuperscript{15} $M_{\odot}$ for galaxy clusters \citep{Bower2004}. 
In this framework, groups are interesting environments for studying the galaxy evolution, since 
they are the place where galaxies spend most of their evolutionary life \citep{Miles2004,Robotham}. 
Galaxy groups are more abundant than galaxy clusters, and most galaxies
in the local Universe, about 55$\%$, are found in group environments \citep{eke2004}. 
The whole scenario of groups or clusters mass assembly is { tricky}, involving many different physical mechanisms.
During the infall of smaller system in larger ones, such as  galaxies in clusters or groups, 
stars can be ripped out from the progenitor galaxies and some galaxies can also be totally destroyed.
In addition, in a dense environment, the mutual interaction among galaxies can build up 
tidal structures such as tails, shells, bridges or tidal dwarf galaxies of baryonic matter stripped away 
from the outer parts of the interacting galaxies.
During the assembly process, the stripped baryons can be in part reabsorbed by the parent galaxies and can in part merge into the 
Brightest Galaxy of the Group/Cluster (BGG, BCG) or in other massive members. The remaining part of the stripped material gets trapped in the group/cluster potential well and forms the extended, 
diffuse and very faint ($\mu_g \geq 26.5$~mag/arcsec$^2$) component known as intragroup or intracluster diffuse light (IGL, ICL)
\citep{Rudick2009,Cui_2013,Mihos2015,Contini2014,Montes2014,JimnezTeja_2019, Montes2018,Pillepich2018,DeMaio2018,henden2019baryon,Contini2019,Montes2019,deMaio2020}. 
In groups, the dynamical timescale is longer and IGL imprint  (i.e. the stripped material) remains for many Gyrs \citep{Watkins2014}. 
According to this scenario, the IGL, the precursor of the ICL, 
can be considered as a fossil record of past interaction events \citep[e.g.][]{Adami2005A&A...429...39A,Mihos2015,Canas2020}. 
The IGL (ICL) inhabits the intragroup (intracluster) space between the galaxies and it is bound only to the group's (cluster’s) potential well, 
and to none of the galaxies that build up the systems. 
Based on this assumption, ICL has also been used as a luminous 
tracer for dark matter \citep[e.g.][]{Montes2018,Montes2019}.
The amount of IGL provides clear information on the dynamical evolution of the structures also because it traces the accretion history of the system and the tidal encounters \citep{Merritt1984ApJ...276...26M}.
Numerical simulations suggest that a larger amount of IGL (ICL) is expected when numerous interactions and encounters between the galaxies of the group (clusters) happen \citep{Murante2007,Purcell2007,Conroy2007ApJ...668..826C,Puchwein2010,Rudick2011,Contini2014,Cooper2015}.
Recent studies revealed a fraction of diffuse light in a range of 10 - 50 $\%$ of the total cluster light, despite these estimates are affected by large uncertainties (see e.g. \citet[][]{Montes_igl2019arXiv191201616M} for a recent literaure review on the history of the ICL).
The observational studies on the distribution of ICL and its physical properties 
(e.g. amount, color) are fundamental to set constrains for numerical simulations \citep{Lin2004ApJ...617..879L,Arnaboldi2010HiA....15...97A,Contini2014,Pillepich2018,contini2020ApJ...901..128C}, and to reduce the discrepancy between observations and cosmological simulations on the baryon component of the universe \citep{Buote2016ApJ...826..146B}. 
An aspect that is still much debated in literature, both from a theoretical and observational side, 
regards how the ICL fraction relates to the virial mass ($M_{vir}$) of the host environment \citep[][]{montes2019intracluster, Contini2021Galax...9...60C}.
From a theoretical point of view, \citet{Lin_2004}, \citet{Murante2007}, \citet{Purcell2007} and \citet{Henden2019MNRAS.489.2439H} found a slight increase in the ICL fraction in systems with higher $M_{vir}$. 
In particular, \citet{Purcell2007} estimated that the ICL fraction increased from 20$\%$, in halos with $M_{vir} = 10^{13} $~M$_{\odot}$, to $30\%$ in halos with $M_{vir} = 10^{15} $~M$_{\odot}$.
In contrast, \citet{Sommer-Larsen2006,Monaco_2006,Dolag2010MNRAS.405.1544D,Henriques_2010,Rudick2011,Contini2014} 
found a fraction of ICL in a range between 20\% and 40\% for all values of $M_{vir}\sim 10^{13}-10^{15}$~M$_{\odot}$. 
\citet{Cui_2013} indicated a slight decrease of the ICL fraction with increasing $M_{vir}$.
On the observational side, \citet{Sampaio-Santos2021} for a sample of 528 clusters at 0.2 $\leq$ z $\leq$ 0.35, 
didn't find an increasing dependence between the amount of ICL and $M_{vir}$. Also from observations in the local universe ($z\leq 0.05$) it seems there is no evident relationship between the ICL fraction and 
the $M_{vir}$ \citep[e.g.][and reference therein]{RAGUSA2021}. A large scatter is observed, where 
high ICL amount ($\sim$ 30 - 45$\%$) occurs both in groups with and in massive 
clusters of galaxies, like Fornax and Antlia (Ragusa et al. 2022 in preparation), as well as low ICL amount ($\sim$ 5 - 10$\%$) results both in groups and in massive clusters of galaxies, like Coma and Abell85 \citep{Jim_nez_Teja_2019,montes2021buildup}.
This scatter might depend on the different concentration (i.e. compact groups vs loose groups) and on the 
different formation time of the particular halo in which the ICL lies \citep[see the review of ][for more details]{Contini2021Galax...9...60C}.
Indeed, more dynamically evolved systems, should host a larger fraction of ICL, as stellar stripping and mergers are more frequent \citep[e.g.][]{Murante2007,Rudick2011,Martel2012,Contini2014}.
In the last two decades, thanks to enhancement of new deep photometric surveys, a great contribution 
was provided to the study of low-surface brightness (LSB) structures in galaxy groups/clusters out to the intragroup-intracluster space, and in particular to the detection of the ICL \citep[e.g.][]{Ferrarese2012,vanDokkum2014,Duc2015,Fliri2016MNRAS.456.1359F,Munoz2015,Merritt2016,Watkins2016ApJ...826...59W, Trujillo2016ApJ...823..123T,Mihos2017,Huang_2018,Montes2019,Zhang2019ApJ...874..165Z,deMaio2020,montes2021buildup,Delgado21}. 
The {\it VST Early-type GAlaxy Survey} \citep{Capaccioli2015} takes its place in this context as a pivotal survey 
with an increasingly important role in the development of the deep and wide-field photometry. 
VEGAS\footnote{Visit the website \url{http://www.na.astro.it/vegas/VEGAS/Welcome.html}} is a multi-band 
$u$,$g$,$r$,$i$ imaging survey, able to map the surface brightness of galaxy down to the azimuthally averaged surface 
brightness $\mu_g$ $\sim$ 31 mag arcsec\textsuperscript{-2} and out to $\sim$ 15 effective radii ($R_e$) \citep[see][and references therein]{Iodice2017a,Spavone2018,Cattapan2019,Iodice2019,Iodice_2020,Spavone_2020,Raj_2020,RAGUSA2021}. 
In this work we present new deep data of the LEO I pair, NGC~3379-NGC~3384, as part of the VEGAS sample. 
The new multi-band images represent the deepest data available for this target.
These allowed us to map the distribution of the IGL and GCs, and to study the physical properties of these tracers in the intra-group region.
This work is organized as follows. In Sec.~\ref{subsec:LEOpair} we introduce the target and the previous studies available in the literature. In Sec.~\ref{sec:obs} we present the observations and the data reduction.
In Sec.~\ref{sec:analysis} we describe in detail the method used for the data analysis (i.e. the surface photometry and GCs detection). In Sec. ~\ref{sec:results} we illustrate the results on the intra-group baryons (diffuse light and GCs) in the LEO I pair NGC~3379-NGC~3384 and compare our results with the previous studies, both on the observational and theoretical side. 
Finally, in Sec.~\ref{sec:conc} we discuss our results and draw the main conclusions of this work.

 \begin{figure}[h!]
 \begin{center}
 \includegraphics[width=17cm]{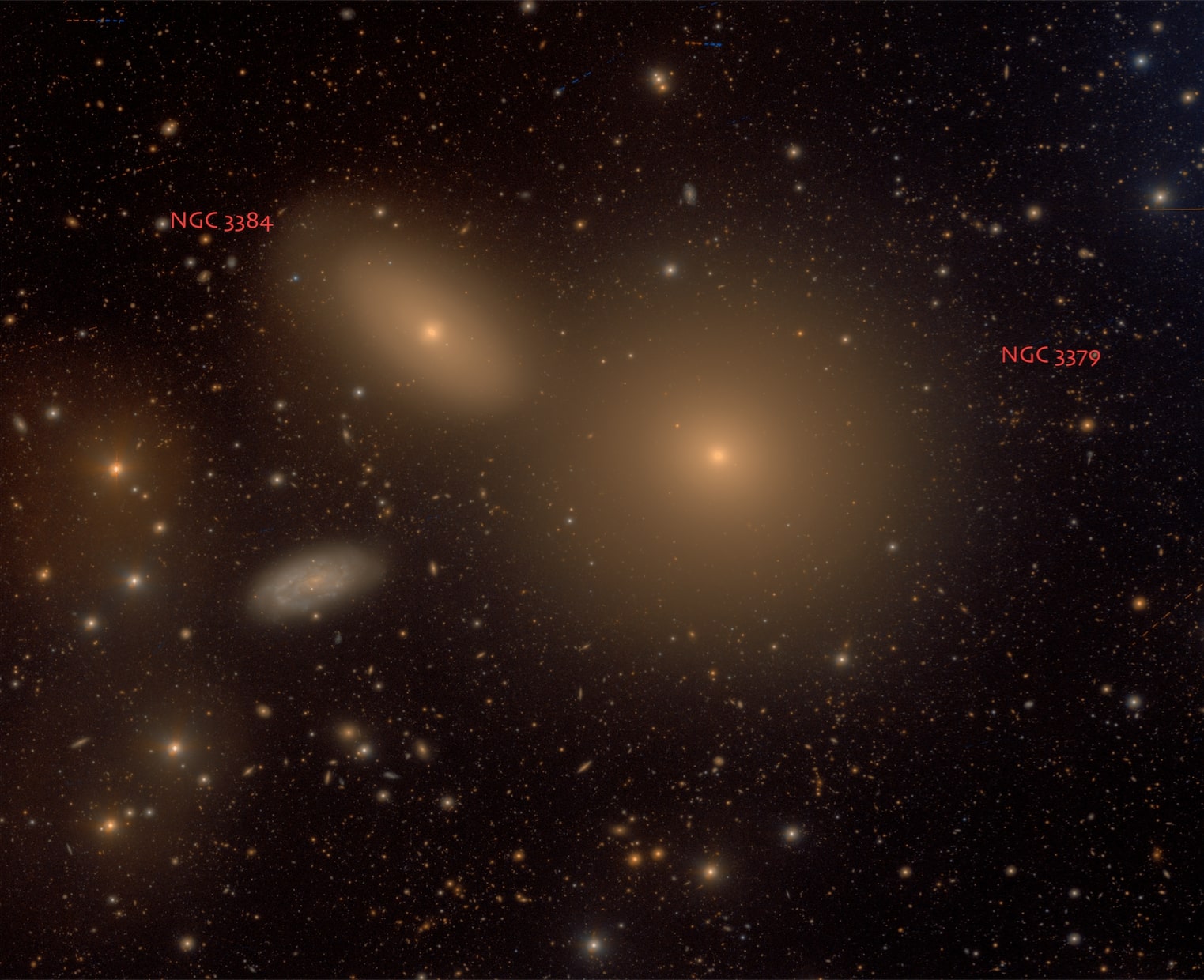}   \end{center}
\caption{\ bf Color composite ($gri$) VST image of the central regions of LEO I pair. 
The image size is $40 \times 37 $~arcmin.  
North is up and East is to the left. The two brightest group members are labelled in red on the image. }\label{fig:composite}
 \end{figure}

\subsection{The LEO I pair NGC~3379-NGC~3384}\label{subsec:LEOpair}

The galaxies NGC~3379-NGC~3384 are members of the LEO I group.
This is a very loose group composed, at least, of 11 bright galaxies where seven out of the total, 
including NGC~3379 (BGG) and NGC~3384, are associated with the subgroup NGC~3368 (M96) centered on the BGG. 
The other four galaxies are members of the {\it LEO triplet} \citep{deVaucouleurs1975gaun.book..557D}. 
In addition to both early and late morphological types of galaxies (ETGs, LTGs), a large number of dwarf galaxies have been revealed by \citet{Muller2018A&A...615A.105M}.
There is extensive literature on the LEO I group, in particular about the presence of a fragmented ring of neutral hydrogen surrounding the pair NGC~3379-NGC~3384, with a radius of about 100 kpc. 
The debate on the history of the formation and evolution of this HI ring is still open and controversial.
To date, no optical counterpart has been detected, except for three far and near-ultraviolet small sources \citep{Thilker2009Natur.457..990T} residing in the high density region of the HI map \citep{Schneider1986AJ.....91...13S}. 
\citet{Schneider1985ApJ...288L..33S} studied the kinematics of the HI gas in the ring, providing an elliptical orbit centered in the light-weighted barycenter between NGC~3379 and NGC~3384. According to this kinematic model, the rotation period of the structure should be $\sim 4$ Gyr \citep{Schneider1985ApJ...288L..33S}. These findings,
combined with the lack of optical counterpart and star formation associated with the ring, set at 4 Gyr the lower limit of its age \citep{Pierce1985AJ.....90..450P,Schneider1985ApJ...288L..33S,Donahue1995ApJ...450L..45D}.
\citet{Sil'chenko2003ApJ...591..185S} found that the ages and kinematics of both the circumnuclear stellar and gaseous disks in NGC~3384, NGC~3379 and M96 are consistent with a formation age on the Gyr scale.
The primordial formation of the HI ring, however, is hindered by stability problems over large time scales, a fundamental inconsistency which suggests that a close encounter between disk galaxies is the most likely scenario for the formation of the ring. They might be M96 and NGC~3384, since the S0 NGC~3384 is the only galaxy of the group having 
the disk major axis aligned with the ring. NGC~3384 could have been transformed into a S0 galaxy after that encounter \citep{Corbelli2021ApJ...908L..39C}. 
According to this hypothesis, \citet{Dansac2010ApJ...717L.143M} reproduced all the most important features of the ring,
such as the shape, the rotation and the lack of the optical counterpart brighter than $\sim$ 28 mag/arcsec$^2$.
Although the formation picture seems confirmed, the absence of any optical counterpart of this structure still remains an open issue. \citet{Watkins2014}, with the deepest photometry present in the literature on this target, did not find any optical counterpart of the ring, down to $\sim$ 30 mag/arcsec$^2$ in the $B$ band.
The authors also suggested that there is no IGL detected down to the SB limit they can explore. 
According with simulations, in the galaxy groups the expected IGL fraction should vary from 12\% up to 45$\%$  \citep{Sommer-Larsen2006,Rudick2006}.
Since the halo mass of the LEO I group falls slightly below the resolution limit of the simulations, 
\citet{Watkins2014} inferred that in this mass range the IGL could contribute significantly less than the larger masses.

The most recent estimate of the IGL amount in the LEO I group is  provided by \citet{Hartke2020}, using photometric and 
kinematic data of the planetary nebulae (PNe). They derived $\sim$ 4$\%$ as a lower limit for the amount of IGL, 
updating the previous estimate of 1.6$\%$ predicted by \citet{Castro2003A&A...405..803C}, also using the PNe 
as discrete tracers for the diffuse intra-group light.
Other studies of the LEO I pair were conducted by \citet{Harris2007}, who found evidence of a transition from a metal rich stellar population to a metal poor stellar population towards the outside of the NGC~3379 stellar halo. Indeed \citet{Harris2007} found a range of  [Fe/H] $\sim -1.3 to -1.5 $for the NGC~3379 stellar halo, more metal poor than the inner regions.
This result was confirmed later by \citet{Lee2016ApJ...822...70L}, who  supported the existence of two different sub-populations of GCs both in colour and metallicity in the halo of NGC~3379: a dominant, red, metal rich
population and a much fainter, blue, metal poor one.
The origin of the stellar halo is still much debated. 
The most accredited scenario proposed by \citet{Lee2016ApJ...822...70L}, foresees two phases: the 
red and  metal rich halo was formed through in situ star formation and/or accretion via major mergers of massive progenitors, while the blue and metal poor halo occurred through dissipationless mergers and accretion events.

\section{Deep VEGAS images of the LEO I pair}\label{sec:obs}

The LEO I pair is a target of VEGAS, a  multi-band $u$,$g$,$r$ and $i$ imaging survey, carried out with the Very Large  Telescope Survey Telescope (VST). The VST is a 2.6 m wide field optical telescope \citep{Schipani2012} equipped with OmegaCAM, a one square degree camera with a resolution of 0.21 arcsec~pixel$^{-1}$.
Fig.~\ref{fig:composite} shows the sky-subtracted color composite $g$, $r$ and $i$ VST image obtained for LEO I pair. The bright spiral galaxy NGC~3389, also visible in the field-of-view (FOV) SW of NGC~3379, 
is a background source and it is not part of the group.
The data used in this paper were acquired in service mode (run IDs: 096.B-0582(B), 097.B-0806(A), 090.B-0414(D)), in clear conditions during the dark time, with an average seeing of  FWHM$\sim$ 1.0 arcsec in the $u$ band, FWHM $\sim$ 1.00 arcsec in the $g$ band, FWHM $\sim$ 0.81 arcsec in the $r$ band and FWHM $\sim$ 0.95 arcsec in the $i$ band. The total integration times are: 2.04 hours in $u$ band, 2.13 hours in $g$ band, 2.06 hours in $r$ band and 0.35 hours in $i$ band.
The surface brightness depths at $5 \sigma$ over the average seeing area listed above are about $\mu_u=24$~mag/arcsec$^2$, $\mu_g=25$~mag/arcsec$^2$, $\mu_r=24.5$~mag/arcsec$^2$ 
\citep{Iodice2021arXiv210204950I}.
The point source depth at 80$\%$ completeness level is 24.38~mag, 25.04~mag, 24.85~mag, 23.44~mag, in the $u$, $g$, $r$ and $i$ band, respectively.

\begin{figure}[h!]
\begin{center}
\includegraphics[width=19cm]{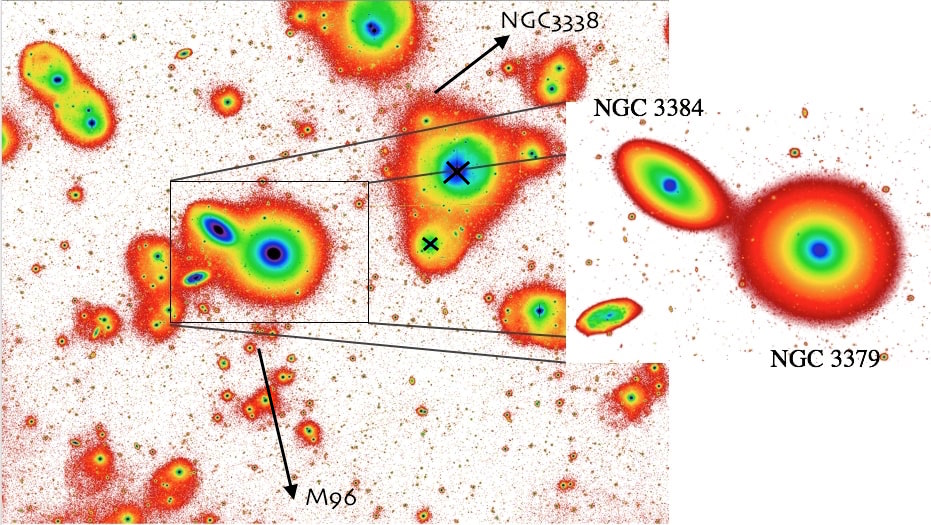}    
\end{center}
\caption{ Deep VST image, of 1.34 x 1.05 square degree, in the $r$ band centred on NGC~3379, used for the analysis described in this paper. 
North is up and East is to the left. The black crosses mark the center of the two brighter stars in the field. The two black arrows represent the direction of M96 and NGC~3338, being these two other brighter galaxies of the Leo I group. The two galaxies that compose the pair are indicated in the enlarged panel shown on the right.}
\label{fig:smooting}
\end{figure}
In the LSB regime, a fundamental step is the estimate of the sky background contribution. 
To this aim, the step-dither observing sequence is adopted.
It consists of a cycle of short exposures (150 sec), both on the target ({\it science frame}) and the sky ({\it offset frame}). The latter is used to estimate the sky background contribution. This turned to be the best strategy to analyze the fields with very bright galaxies, having a very extended envelope \citep[see e.g.][]{Iodice2016,Spavone2017a,Iodice_2019}.
In this particular approach, the offset frame is taken as close as possible, in space and time, to the science frame (offset by $ \pm 1$°), but far enough from the central bright galaxies, to avoid taking into account the diffuse light of the extended envelope of the galaxies and IGL. 
The data reduction was carried out by using the $VST-Tube$ pipeline \citep{Grado2012}, which is described in detail
by \citet{Capaccioli2015} and \citet{Spavone2017a}.
The VST mosaic covers an area of 3.65 square degrees centered on the LEO I pair.
In  Fig.~\ref{fig:smooting} we present the deep VST image in $r$ band, of 1.34 $\times$ 1.05 square degree, centred on the LEO I pair. 
The main properties of the two galaxies are listed in Tab. \ref{tab:parameterNED}.
In this paper we adopted for LEO I pair  the same average distance of the BGG, 10.23$\pm$ 0.08 Mpc \citep{Lee2016ApJ...822...70L}, 
so 1 arcsec corresponds to $\sim$ 0.05 kpc. 
The extinction-corrected \citep{Schlafly_2011} magnitudes through-out the paper are in the AB system.

\section{Data analysis: Surface photometry and GCs detection}\label{sec:analysis}

The science goals of this study are to map the intra-group baryons made by diffuse light and GC systems. In particular, we aim at estimating the IGL contribution to the total 
luminosity of the group and to provide its $g-r$ color. 
To this aim the crucial task is to separate the contribution of the IGL from that of the scattered light by the foreground/background sources and by the light belonging to the group members.
The main steps we performed for the data analysis are briefly summarised below and discussed in detail 
in the next subsections. They have been tested and successfully applied to all VST images \citep[see][and references therein]{Iodice2017a,Spavone2018,Cattapan2019,Iodice2019,Iodice_2020,Spavone_2020,Raj_2020,RAGUSA2021}, 
to study the LSB regime, in particular to detect the IGL.
For the estimate and analysis of the IGL we use the $g$ and $r$ bands only, since these are
 the deepest images with the most efficient filters of OmegaCAM.
On the sky-subtracted mosaic we proceed as follow:

\begin{itemize}
    \item the contamination from the foreground brightest stars in the field are removed by subtracting their models from the image (see Sec. \ref{subsubsec:stars});
    \item the limiting radius of the photometry (R$_{lim}$) and the residual background fluctuations (see Sec. \ref{subsubsec:Rlim}) are estimated;
    \item the isophote fitting of the brightest group members out to R$_{lim}$ (see Sec. \ref{subsubsec:Rlim}) are performed to obtain the azimuthally-averaged surface brightness profiles and shape parameters;
    \item the brightest group members have been modelled and subtracted out to their transition radius (R$_{tr}$) (see Sec. \ref{subsubsec:fit1d}).
\end{itemize}
The tools and methods adopted in each step are described in the following subsections.

\subsection{Scattered light from the bright stars}\label{subsubsec:stars}

To account for the scattered light from the two bright foreground stars located North-West in projection 
of the BGG (R.A. 10:46:19.207 Decl. +12:44:47.19 with $m_B$= 9.42 mag, and R.A. 10:47:00.015 Decl. +13:01:37.01 with $m_B$= 10.49 mag, see Fig. \ref{fig:smooting}), we have modelled their light distribution and 
subtracted it from the parent image.
To this aim we have derived a 2-dimensional (2D) fit of the light distribution (using the IRAF\footnote{IRAF (Image Reduction and Analysis Facility) is distributed by the National Optical Astronomy Observatories, which is operated by the Associated Universities for Research in Astronomy, Inc. under cooperative agreement with the National Science Foundation} task ELLIPSE \citep{Jedrzejewski1987}).
Fit is made for both sources assuming a circular light distribution 
out to the edge of the frames ($\sim$ 60 arcmin, see Fig. \ref{fig:smooting}), 
keeping all parameters fixed to their initial values, i.e. center ($x_0$,$y_0$),
ellipticity ($\epsilon$ = 0.05) and position angle (P.A. = 0).
Before performing the isophote fit of the stars, we accurately masked \footnote{All the masks of this work were created by using the IRAF task MSKREGIONS.} all the foreground/background sources. The core of the group is also masked 
out to $\sim$ 20 arcmin ($\sim$ 19 R$_{eff}$) of the BGG \citep{Capaccioli1990}.
Using the IRAF task BMODEL, we built up the 2D models of the stars and subtracted them from both $g$ and $r$ parent images. 

\subsection{Estimate of R$_{lim}$ and residual background fluctuations}\label{subsubsec:Rlim}

Using the same method described by \citet{RAGUSA2021}, the star-subtracted images are used to estimate
the R$_{lim}$ and the average value of the residual background level\footnote{The “residual background” 
in the sky-subtracted images are the deviations from the background with respect to the average sky frame 
obtained by the empty fields close to the target.}.
Since images are sky-subtracted the residual background level is close to zero. This value (and its RMS
fluctuation) has been taken into account when computing 
the galaxies' surface brightness and the relative uncertainties, i.e. all surface brightness profiles are corrected for such a residual value of the background.
The R$_{lim}$ corresponds to the outermost semi-major axis derived 
by the isophote fitting, with respect to the center of the target, where the galaxy's light blends into the average residual background level. Beyond R$_{lim}$ the residual background fluctuations are almost constant. Since the two galaxies in the LEO I pair are close in projection and mostly located at the image centre, the isophote fit is performed by adopting the BGG centre and, therefore, the  $R_{lim}$ is the same for both galaxies. 
As explained in \citet{RAGUSA2021}, by using IRAF task ELLIPSE, light is fitted in circular annuli (i.e. the ellipticity and PA are fixed to zero), with constant step, 
out to the edge of the images. All the foreground/ background sources and the companion NGC~3384 galaxy are accurately masked. We found that R$_{lim}\sim 20$ arcmin in both $g$ and $r$ bands. 
For $R\geq R_{lim}$, the residual background levels are I$_g$ $\sim$ - 0.52 $\pm$ 0.02 ADU and 
I$_r$ $\sim$ - 0.90 $\pm$ 0.03 ADU in the $g$ and $r$ band, respectively.

\subsection{Isophote fitting}\label{subsubsec:isoph}

In order to derive the amount and physical properties of the IGL, we need to account for
the contribution to the light from NGC~3379 and NGC~3384. 

\begin{figure}[h!]
\centering
\resizebox{0.9\textwidth}{0.7\textheight}{\includegraphics{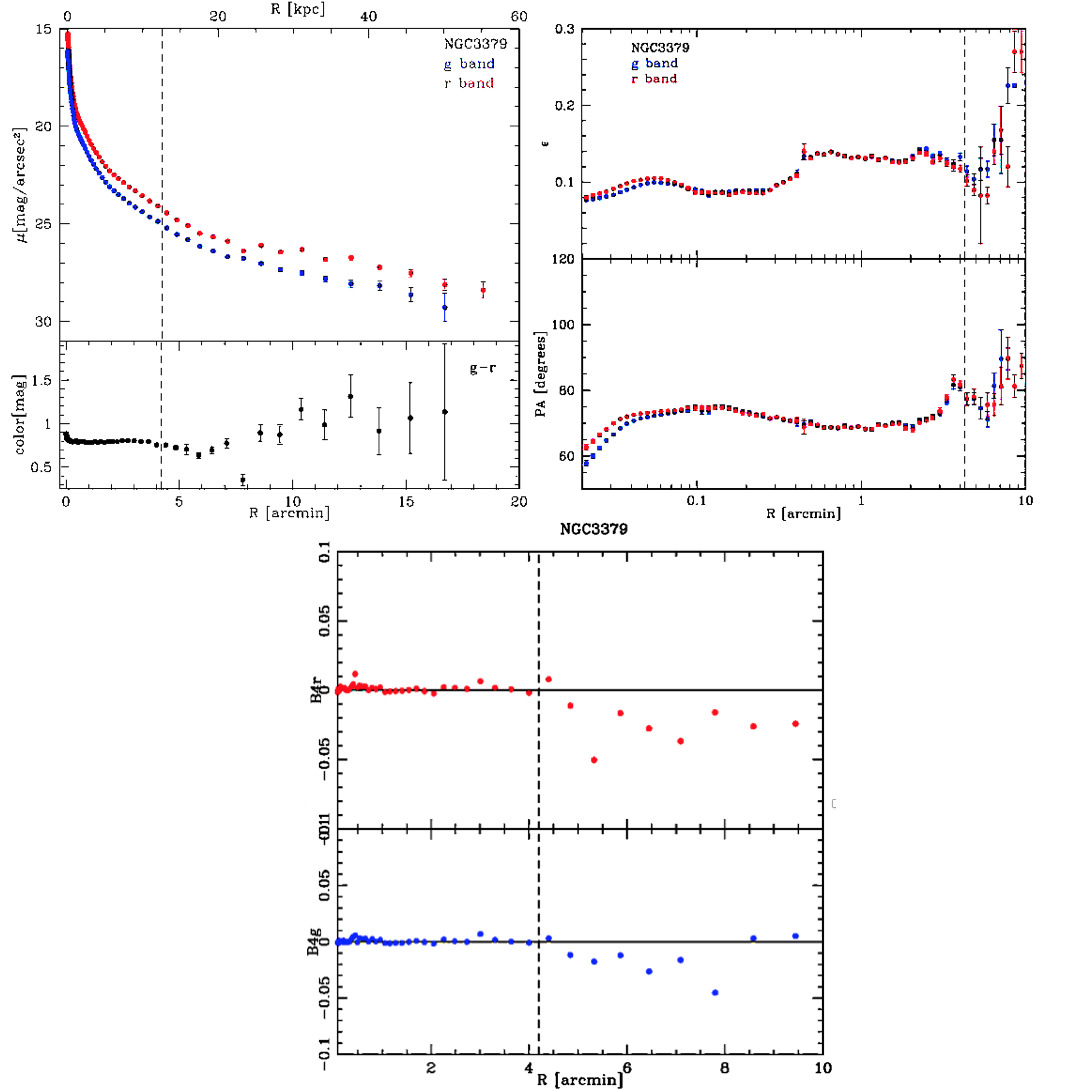}}
\caption{{\it Top left panels:} azimuthally-averaged surface brightness profiles of NGC~3379 in VST $g$ (blue) and $r$ (red) bands (top panel) and $g-r$ color profile of NGC~3379 (bottom panel). {\it Top right panel:} azimuthally-averaged ellipticity (top panel) and position angle (bottom panel) profiles of NGC~3379 in VST $g$ (blue) and $r$ (red) bands. The black dashed lines correspond to the R$_{tr}$ (see Sec. \ref{subsubsec:fit1d}). {\it  Bottom panels:} radial profiles the B4 Fourier coefficient as a function of semi-major axis in VST $r$ (top panel) and $g$ (bottom panel) bands for NGC~3379. The vertical dashed line corresponds to R$_{tr}$, beyond which the isophotes show a boxy shape (see details in the text in Sec. \ref{subsubsec:fit1d}.}\label{fig:ellipseNGC3379}
\end{figure}
To this aim, for both galaxies, we derived the azimuthally-averaged surface brightness profiles and then we performed the 1-dimensional multi-component fit to set the scales of the contribution to the light from the galaxy and IGL (Sec.~\ref{subsubsec:fit1d}).
Firstly, we fitted the isophotes of the BGG, NGC~3379, out to R$_{lim}$. We built a detailed mask on the sky-subtracted and star-removed images, of the foreground/background sources and of NGC~3384, by using IRAF task MSKREGIONS.
The fit of the isophotes was performed using the IRAF task ELLIPSE, with all 
parameters left free (i.e. center, ellipticity and P.A.), over elliptical annuli, by applying a median sampling and k-sigma clipping algorithm. 
The combination of median sampling and k-sigma clipping algorithm has been shown to perform at best 
the fit of the isophotes, rejecting deviant sample points at each annulus.
We derived the azimuthally averaged surface brightness, color and geometrical parameters profiles in each band.

\begin{figure}[h!]
\centering
\resizebox{.9\textwidth}{.7\textheight}{\includegraphics{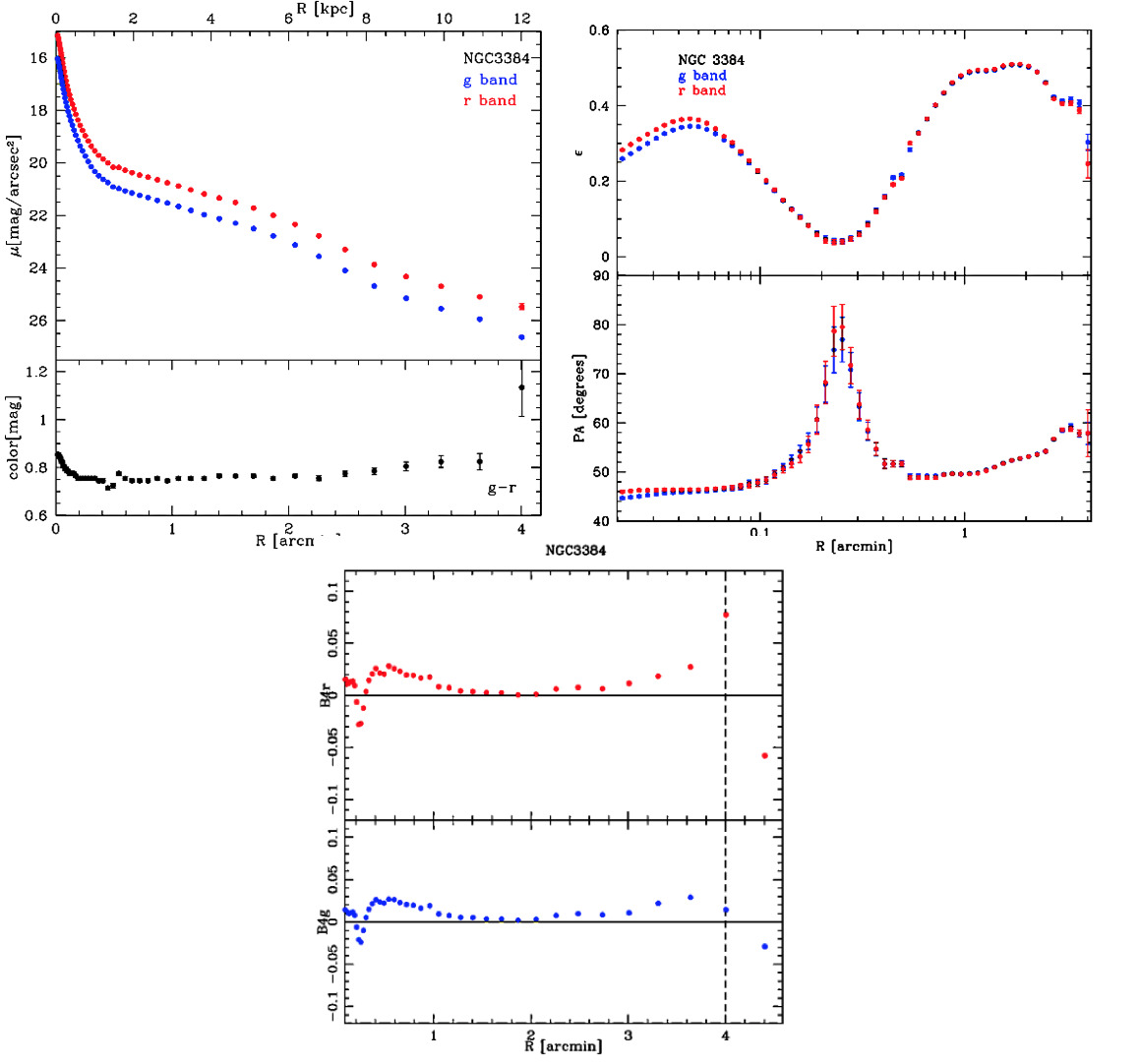}}
\caption{{\it Top left panels:} azimuthally-averaged surface brightness profiles of NGC~3384 in VST $g$ (blue) and $r$ (red) bands (top panel) and $g-r$ color profile of NGC~3384 (bottom panel). {\it Top right panels:} azimuthally-averaged ellipticity (top panel)and position angle (bottom panel) profiles of NGC~3384 in VST $g$ (blue) and $r$ (red) bands.
{\it Bottom panels:} radial profiles the B4 Fourier coefficient as a function of semi-major axis in VST $r$ (top panel) and $g$ (bottom panel) bands for NGC~3384. In this case also the coefficient B4 gives an indication of the lack of a evident accreted component (see details in the text in Sec. \ref{subsubsec:fit1d}).}\label{fig:ellipseNGC3384}
\end{figure}
The total uncertainty on the surface brightness profile takes into account the 
uncertainties on the photometric calibration (i.e. on the zero points, zp$_g$= 30.000 $\pm$ 0.005 mag and zp$_r$= 30.000 $\pm$ 0.004 mag) and the RMS in the background fluctuations\footnote{ Calculated with the following formula: 
$err= \sqrt{(2.5/(adu \times \ln(10)))^2 \times ((err_{adu}+err_{sky})^2) +  err_{zp}^2}$, where
N is the  number of pixels used
in the fit, $err_{sky}$ is the rms on the sky background, $err_{zp}$
is the error on the photometric calibration, adu is the analog digital unit and $err_{adu}=\sqrt{adu/N-1}$ \citep{Capaccioli2015,Seigar2007}.}.
From the isophote fit, we derived the 2D model (using IRAF task BMODEL) of the light distribution for NGC~3379 up to R$_{lim}$ in each band, and subtracted it from the parent image.
On the residual images, we performed the isophote fit of the other galaxy of the pair, NGC~3384, 
up to R$_{lim}$ in each band, following the same procedure used for NGC~3379.
In the top left panels of the Fig.\ref{fig:ellipseNGC3379} and of the Fig.\ref{fig:ellipseNGC3384} we show the azimuthally-averaged surface brightness and $g$ - $r$ color profiles of the two galaxies, NGC~3379 and NGC~3384 respectively, as a function of isophote semi major-axis.
We are able to map the surface brightness profile of the BGG NGC~3379 down to $\sim$ 30 mag arcsec\textsuperscript{-2} in the $g$ band and $\sim$ 29 mag arcsec\textsuperscript{-2} in the $r$ band. 
We also provide the reliable $g-r$ color profile of the galaxy out to $\sim$ 17 arcmin 
(i.e. $\sim$ 50 kpc) from the galaxy centre.
The lower part of the top left panel
in Fig. \ref{fig:ellipseNGC3379} shows that the $g-r$ color profile changes the trend at a radius of $\sim$ 4 arcmin (i.e. $\sim$ 12 kpc). 
Within this radius the color is almost constant, whit a value $g-r \sim 0.8$ mag. At larger distances, the color profile has a negative gradient with a $g-r$ value ranging from $\sim$ 0.8 mag, at 4 arcmin ($\sim$ 12 kpc), to $\sim$ 0.6 mag, at 6 arcmin ($\sim$ 18 kpc). 
In the top right panels of the Fig.\ref{fig:ellipseNGC3379} and the Fig.\ref{fig:ellipseNGC3384} we show the azimuthally-averaged ellipticity and position angle profiles of the two galaxies NGC~3379 and NGC~3384, 
respectively, as a function of isophote semi major-axis.
A break in the ellipticity and PA radial profiles is also evident, at the same radius ($\sim$ 6 arcmin) in which the $g-r$ color profile of NGC~3379 becomes shallower. 
From the isophote fit we obtained the growth curve, used to compute the total magnitude m$_{tot}$ for NGC~3384 
and NGC~3379 in both $g$ and $r$ bands.
From the extinction-corrected magnitudes (see Tab.~\ref{tab:parameterfit}) we estimated: {\it i)} the average color $g-r$ = 0.76 mag and the total luminosity of NGC~3384, which is $L_g= 1.60\times 10^{10} L_{\odot}$ and $L_r= 1.71 \times 10^{10} L_{\odot}$; {\it ii)} the average color $g-r$ = 0.82 mag and the luminosity of NGC~3379, out to its transition radius (i.e. the radius beyond which the IGL plus diffuse stellar envelope of the galaxy start to dominate, see Sec.~\ref{subsubsec:fit1d}), which is $L_g= 4.44\times 10^{10} L_{\odot}$ and $L_r= 4.95 \times 10^{10} L_{\odot}$.

\subsection{1-dimensional fit of the light distribution: the contribution of the IGL}\label{subsubsec:fit1d}

\begin{figure}[h!]
\centering
\resizebox{1.\textwidth}{.43\textheight}{\includegraphics{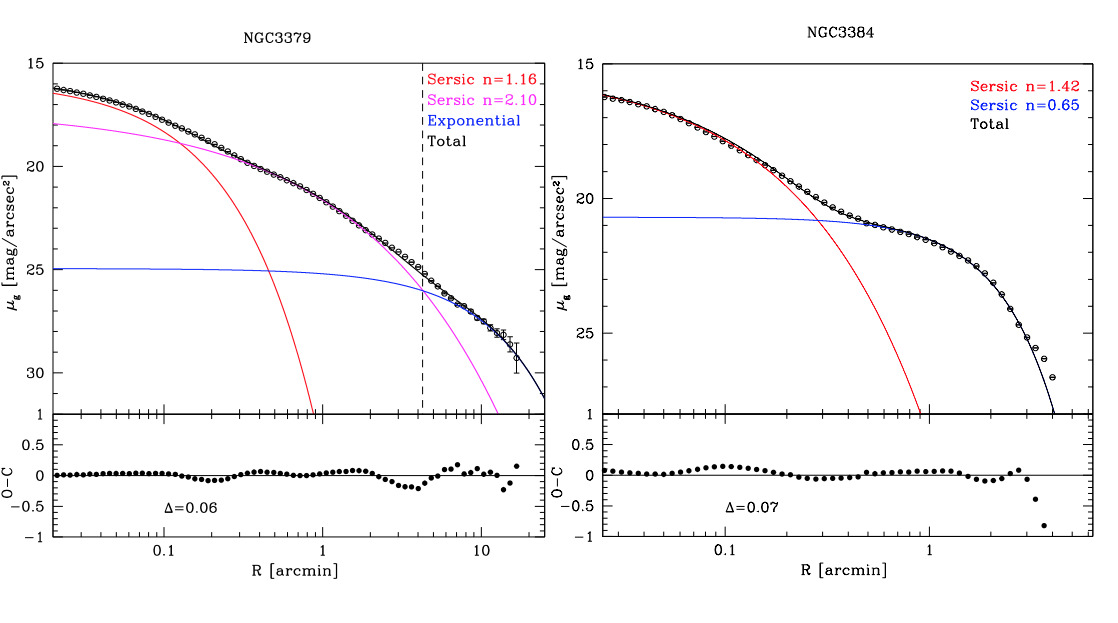}}
\caption{{\it Left panel:} (Top panel) Three components model of the azimuthally averaged surface brightness profile of NGC~3379 out to $\sim$ 57 kpc in $g$ band. The red and magenta lines indicate a fit to the inner region with a two Sérsic profiles. The blue line indicates a fit to the outer diffuse component (IGL) and the black line indicates the sum of the components in each fit. The vertical dashed line shows the estimated value for R$_{tr}$ ($\sim$ 12.5 kpc).(Bottom panel) $\Delta$rms scatter of the data minus the model (see text for details). {\it Right panel:} (Top panel) Two components model of the azimuthally averaged surface brightness profile of NGC~3384 out to $\sim$ 12 kpc in $g$ band. The red and blue lines indicate a fit to the inner region with a two Sérsic profiles. The black line indicates the sum of the components in each fit. 
(Bottom panel) $\Delta$rms scatter of the data minus the model (see text for details). }\label{fig:1Dfit}
\end{figure}

The massive galaxies at the center of clusters or groups are made by a very bright central component, 
fitted with one or two S{\'e}rsic law and a diffuse and very extended envelope, which is 
well reproduced by an exponential law
\citep[see e.g.][]{Seigar2007,Donzelli2011,Arnaboldi2012,Huang2013,Cooper2013,Iodice2016, Iodice_2019,Spavone2017b,Spavone_2020,RAGUSA2021}.
The exponential envelope takes into account both the stellar halo of the galaxy and the diffuse light component 
around it (i.e. IGL, ICL).
Based on the deep photometric data, it is currently a challenge to unambiguously separate the extended stellar halo 
of the BGG (or BCG), which is gravitationally bounded to the galaxy, from the diffuse light around it. 
There is more than one photometric analysis used to separate the two components \citep[see][as review]{montes2019intracluster}.
In this work we adopted the same method described in \citet{RAGUSA2021}.
In short, we have derived the transition radius (R$_{tr}$) between the brightest parts of the BGG and the galaxy outskirts by fitting the 1D azimuthally-averaged surface brightness profiles, adopting the fitting procedure 
introduced by \citet{Spavone2017b}, also used in other VEGAS papers \citep[see e.g.][]{Spavone2018,Spavone_2020,Cattapan2019,Iodice2016, Iodice_2020}. 
For the BGG of the pair, NGC~3379 (left panel of the Fig. \ref{fig:1Dfit}), the best fit is reproduced by the 
combination of three components. The inner and brightest regions are well fitted by two S{\'e}rsic laws.
The outskirts show an extended exponential component, for $R\geq 250$~arcsec ($\sim 12.5$~kpc), 
with a central surface brightness of $\mu_0=24.65$~mag/arcsec$^2$ and a scale length $r_h=260$~arcsec 
($\sim 13$~kpc) in the $g$ band.
The transition radius R$_{tr}$, which separates the inner and brighter regions of the galaxy from the fainter outskirts is $\sim$ 4.2 arcmin ($\sim$ 12.5 kpc).
The shape parameters (i.e. ellipticity, P.A. and the 4th coefficient of the series a4 and b4) also
show a different trend for $R\geq R_{tr}$. In particular, the outskirts
become flatter ($\epsilon\sim0.2-0.3$) and a twisting of $\sim 20$~degrees is observed (see top right panels of Fig.~\ref{fig:ellipseNGC3379}), with isophotes being more boxy (see bottom panels of Fig.~\ref{fig:ellipseNGC3379}).
For $R\geq R_{tr}$, the $g-r$ color profile shows redder colors ($\sim1-1.3$ mag) with respect to the central regions that have $g-r\sim0.8$~mag (see Fig.~\ref{fig:ellipseNGC3379}).
In order to separate the contribution of the bright part of NGC~3379 from the outskirts, we obtained 
the 2D model of the galaxy up to its R$_{tr}$ $\sim$ 4.2 arcmin and 
subtracted it from the stars-subtracted images.
The 2D model is obtained using the IRAF task BMODEL, having has input the outcome of the isophote fit out to the transition radius.
For NGC~3384 (Fig. \ref{fig:1Dfit}, right panel) the best fit is reproduced by the combination of two S{\'e}rsic components, which model the brightest regions of the galaxy and an outer disk, for $R\sim$ 3 arcmin. 
The 2D model of NGC~3384 is derived and subtracted from the parent image, where the 
2D model of NGC~3379 was also subtracted. 
The structural parameters obtained by the 1D fit, both for NGC~3379 and NGC~3384, are listed in Tab. \ref{tab:parameterfit}.
The final residual images map the light in the stellar envelopes around galaxies plus the IGL component in the LEO I pair. This is one of the main results of this work
and it is shown in Fig.~\ref{fig:igl_map} for the $g$ band. Here, the light from the background spiral galaxy NGC~3389 has also been modelled and subtracted.

\subsection{Census of the Globular clusters}\label{subsec:GC analysis}

Globular clusters (GCs) in massive galaxies are very old stellar systems, bright and typically fairly numerous. These characteristics, together with the many relevant properties that can be studied (radial and color distribution, luminosity function, etc.) make GCs a useful fossil record of the galaxy and its host environment \citep{Harris2001,Cantiello2020}.
Thanks to the large area covered by the VEGAS images, and the number of available passbands, we took advantage of our VST data to study the GC population around the LEO\,I pair.
The analysis is carried out using the same methods described on the previous papers of the Fornax Deep Survey (FDS) and VEGAS surveys \citep[e.g.][]{Cantiello2015,Cantiello2018,Cantiello2020}. Here we briefly summarize the main steps of the GC photometry, selection and analysis.

\subsubsection{GCs photometry}

The photometry of point-like and background extended sources in the frame was obtained running SExtractor \citep{Bertin1996} independently on the $ugr$ and $i$ galaxy-subtracted frames.
The detection parameters (SExtractor's DETECT\_MINAREA, DETECT\_THRESH, SEEING, weighting maps, etc.) were optimized for each passband to improve the detection of the faintest sources. In particular, we visually inspected the central galaxy-subtracted regions to verify that no obvious source is missing, or that surface brightness fluctuations \citep{ts88} are identified as real sources.
We adopted the 8-pixel ($\sim1.7 arcses$) aperture magnitude as our reference. The aperture correction to the 8-pixel magnitudes are derived using the classical curve of growth analysis on isolated and bright stars in the field \citep{Cantiello2007bi}. The values of correction lie in the range of $0.3$-$0.6$ mag (larger corrections for images with wider FWHM). The aperture and extinction corrected catalogs of each individual passbands are then matched using a $1.0~arcsec$ matching radius, taking the $r$ as our reference passband, because of the better image quality. The final catalog of matched sources contains $\sim 23000$ sources, composed by the MW stars, the background galaxies and the GCs in the area.
For each detected source, SExtractor also provides other quantities that we used to identify GCs: Elongation, FWHM, Flux Radius \citep[see ][for a detailed description of these quantities]{Bertin1996}. We also measured the 4 and 6 pixels aperture magnitudes, to derive the concentration index (C.I.) of the sources which we also used to sort out GCs from the full matched catalog \citep{Peng2011ApJ...730...23P}.

\subsubsection{GCs selection}

\begin{figure*}[h!]
\begin{center}
\centering
\resizebox{1.\textwidth}{.4\textheight}{\includegraphics{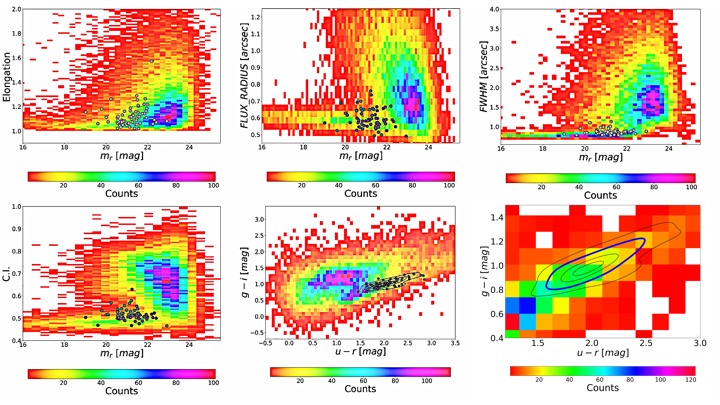}}
\caption{{Upper panels: SExtractor output parameters in the $r$-band for the full sample of detected and $ugri$ matched sources vs corrected $r$-band magnitude $m_r$. Black circles show the GCs from the master catalog. Lower panels: As upper, but concentration index (C.I., left panel) vs $m_r$ and color-color planes (middle and right panel) are shown. In the middle panel, the isodensity contours of GCs from the master catalogs are also shown. In the right panel the contours, and the second isodensity contour (in blue containing $\sim60\%$ of objects of the master catalog) used for the final GCs identification, are overlaid to the 2D-color histogram.
}}\label{fig:selGC}
\end{center}
\end{figure*}

To identify GCs from the final $ugri$ matched photometric catalog, we relied on the observed properties of confirmed GCs from existing studies, and on the known properties of the GCs in the area. 
We built a master GCs catalog, to be used as reference for selecting GCs over the wide LEO\,I area, taking as reference the available catalogs of GCs identified: $i)$ from spectroscopic analysis \citep{Puzia_2004,Bergond_2006}, 64 GCs in total $ii)$ from combined $UBRI$ and $gri$ photometry  \citep{Whitlock_2003,Faifer_2011}, 51 GCs; $iii)$ using X-ray selected sources from catalogs of LMXB \citep{Brassington_2008,Kundu_2007}, 11 GCs. After matching the whole set of targets from the literature, our master catalog contained 67 GCs. 
Table \ref{tab:masterGC} contains the master GCs catalog. Figure \ref{fig:selGC} shows the GCs from the master catalog overlaid with the 2D histograms of measured properties of sources in the full matched catalog.

Adopting the observed ranges of GCs properties from the master catalog, we defined the intervals to be used for identifying bona-fide GCs. The values of FWHM, concentration index, elongation, flux radius, photometric uncertainty and magnitudes adopted are reported in Table \ref{tab:selGC}. These ranges are obtained from the median of the master catalog $\pm 3 \sigma$. GCs also show a universal luminosity function \citep[GCLF;][]{Harris2001}, which we used to define the magnitude range for GCs. The GCLF has a gaussian shape with peak at $M_g\sim-7.5$ mag \citep{Harris2001,Villegas2010}, and width $\sigma_{GCLF}$ that depends on the galaxy mass/luminosity. Using the empirical relations from \citet{Villegas2010}, we estimated  $\sigma_{GCLF}=1.05$ mag in the $g$ band. At the adopted distance of 10.23$\pm$ 0.08 Mpc for LEO\,I, which implies GCLF peak at $m_g\sim22.5$ mag, we select as GCs candidates the objects within $\pm~3\sigma_{GCLF}$ from the GCLF peak: $19.0 \leq m_r~(mag)\leq 26.0$ mag. 
Using the parameter ranges defined as described above, 2397 GCs candidates are pre-identified. We finally inspected the color-color properties of the selected sources to further clean the sample. Figure \ref{fig:selGC} shows the color-color selection procedures adopted: the middle lower panel shows the color-color diagram for the full sample of matched sources, and the isodensity levels of the master catalog; the right lower panel shows the parameter-selected GCs with highlighted the isodensity level we finally adopted to obtain our final catalog of GC candidates (blue contours in the panel), containing a total of 268 candidates \citep[see also][]{Cantiello2020}.
The final list of matched and selected GCs is reported in Table \ref{tab:GC}.
The final GCs catalog still contains unresolved contaminating sources matching with the properties of the GCs population we are interested in. Nevertheless, thanks to the large area covered by our images, and assuming that any population of contaminants is basically uniform over the inspected area, the GCs in the region will be analysed using background subtraction methods (Sect. \ref{subsec:GC}).

\section{Results: The intra-group diffuse light in the LEO I pair  NGC~3379-NGC~3384}\label{sec:results}

\begin{figure}[h!]
\begin{center}

\includegraphics[width=16cm]{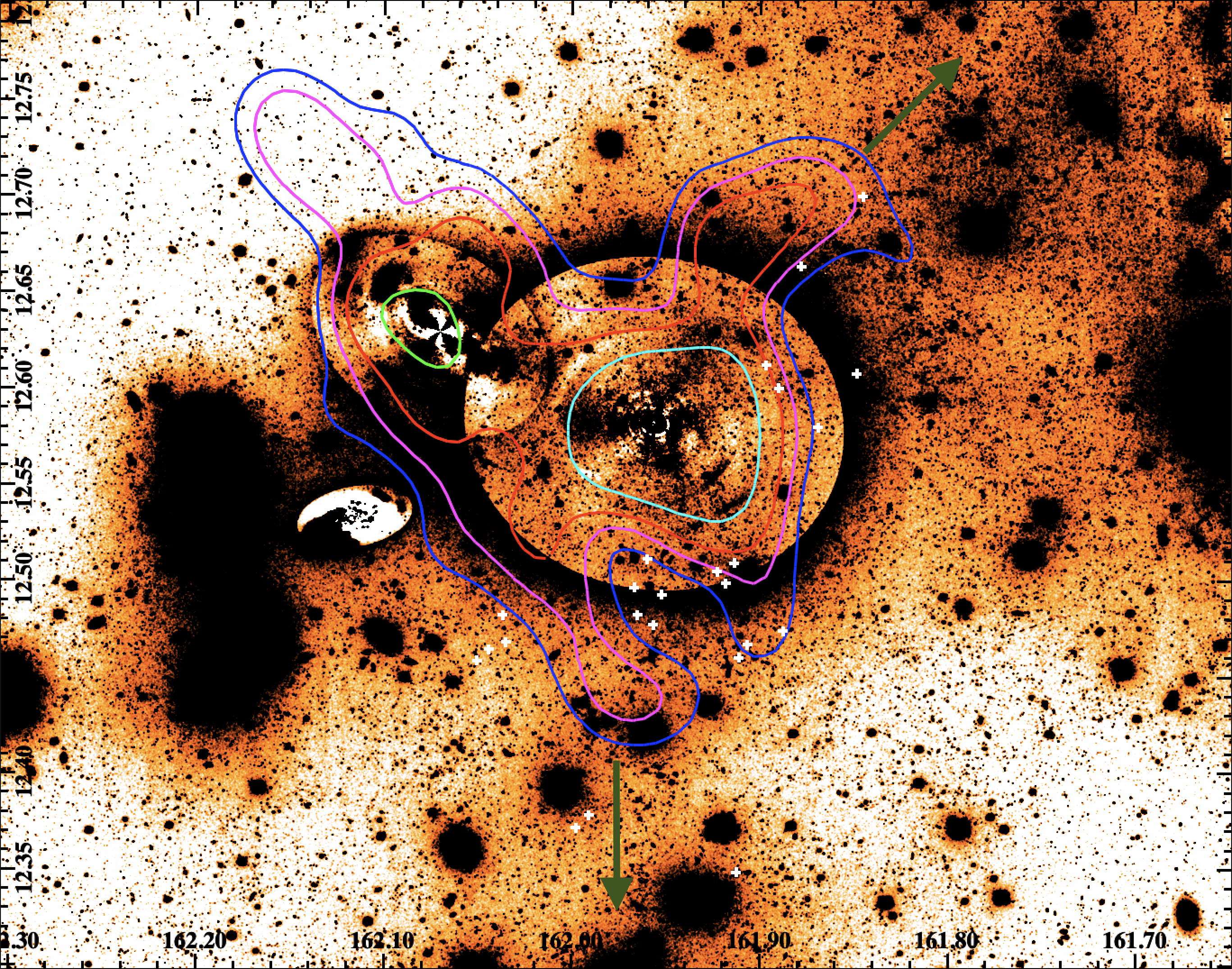}
\caption{Residual image of the LEO I pair in the $g$ band, where the brightest regions of the group members were subtracted from the parent image (see Sec.~ \ref{subsubsec:fit1d}). The image is $0.64 \times 0.50$~degree, inside $R\leq R_{lim}$, where the IGL is computed.
The two green arrows indicate the directions of the two streams that come out of the analysis of the IGL map 
(for details see Sec.~\ref{sec:results}). The solid coloured lines indicate the iso-density contours of the GCs distribution (see Sec.~\ref{subsec:GC}).
The white crosses mark the PNe obtained by \citet{Hartke2020}, which overlap with the two streams.}
\label{fig:igl_map}
\end{center}
\end{figure}

In this section we present the main properties of the IGL in the LEO I pair and the connection with the GCs population in the group,
based on the deep optical images presented in this work. 
In addition, the photometric $R_{tr}$ is compared with the kinematic transition radius found by \citep{Hartke2020} by using PNe.
The distribution of the diffuse intra-group light in the LEO I pair NGC\,3379-NGC\,3384, derived in the $g$ band, is shown in Fig.~\ref{fig:igl_map}. As described in Sec.~\ref{subsubsec:fit1d}, this comes from the sky-subtracted image in the $g$ band, where the brightest regions of the group members have been modelled and subtracted off. Residuals show the contribution of the stellar envelopes around the two brightest galaxies, which is symmetrical distributed around them, plus the diffuse light into the intra-group space. 
Here, two very faint streams are detected, emerging from the BGG. One extends to the South, 
towards the other bright group member of the LEO I group, M96, and the second one is protruding to the North-West, 
towards another bright member of the LEO I group, NGC~3338 (see  Fig.~\ref{fig:igl_map}).
We cannot map the whole extension of this second stream since it partly overlaps with the residuals from the subtraction of the bright star located in this area (see Sec.~\ref{subsubsec:stars}).
To estimate the integrated magnitude of the IGL component in both $g$ and $r$ bands, we have used the residual image shown in Fig.~\ref{fig:igl_map}, where all the foreground and background sources are accurately masked. From this the residual image, using the IRAF task PHOT, we estimated the integrated the flux within $R_{lim}$ ($\sim$ 20 arcmin). The extinction-corrected integrated magnitude is $m_g$= $ 10.72 \pm 0.02$~mag and $m_r$= $ 9.97 \pm 0.03$~mag, in the $g$ and $r$ bands, respectively. These values correspond to the integrated magnitudes of the stellar envelope plus the IGL, which has an average color of $g-r = 0.75 \pm 0.04$ mag. We then obtained the total luminosity of $L_g^{IGL}= 1.22 \times 10^{10} L_{\odot}$ and $L_r^{IGL}= 1.28 \times 10^{10} L_{\odot}$ in the $g$ and $r$ bands, respectively.
The total luminosity of the pair ($L_g^{pair}$) is given by the contribution of the total luminosity of NGC~3379 
($L_{tot}$= $L_{NGC~3379}$), out to its transition radius, the total luminosity of NGC~3384 ($L_{NGC~3384}$) 
and the total luminosity for the galaxy outskirts, i.e. stellar envelope plus the IGL ($L_{IGL}$), derived above. 
This is $L_g^{pair}= 7.27 \times 10^{10} L_{\odot}$ and $L_r^{pair}= 7.94 \times 10^{10} L_{\odot}$ in the $g$ and $r$ bands, respectively.
The fraction of IGL with respect to total luminosity of the pair is $\simeq 17 \pm 2\%$ both in the $g$ and $r$ bands. 
Since the total luminosity of the BGG is $L_g^{NGC~3379} = 4.60\times 10^{10} L_{\odot}$ in the $g$ band and 
$L_r^{NGC~3379} = 5.15\times 10^{10} L_{\odot}$ in the $r$ band, the IGL fraction with respect to the BGG is $\sim 25 \%$ in both the $g$ and $r$ bands.
The IGL fraction of 17$\%$ we obtained in the pair is significantly larger than previous estimate for the LEO I group given by \citet{Watkins2014}. Authors suggested
that the amount of diffuse light they found for the whole LEO I group (including also the two bright galaxies M96 and M95)
is negligible ($\leq$0.01 $\%$) with respect to the total luminosity of the group. 
The main difference between the IGL fraction provided in this work and that from \citet{Watkins2014} is due to a different approach adopted for the IGL estimate. 
In detail, \citet{Watkins2014} did not perform a photometric decomposition of the 
observed profile of NGC3379, modelling the whole light distribution of the galaxy, including the outer stellar envelope out to 45 kpc. Therefore, the residual image 
should include only the fraction of IGL outside the modelled radii. 
Conversely, as addressed in Sec.~\ref{sec:analysis}, we have estimated the total 
amount of the diffuse stellar envelope plus the IGL (i.e. the light beyond the $R_{tr}$, obtained from the 1D decomposition) with respect to the total brightness of the pair 
NGC~3379-NGC~3384, and not with respect to the total luminosity of the Leo I group. 
This approach, widely used in the literature (see references in Sec.~\ref{sec:analysis})
is motivated by the fact that, based on the photometry alone, the contribution to the 
light of the IGL cannot be separated from the light in the stellar envelope, since the two components are completely merged in the galaxy outskirts. 
Such a kind of separation is possible using the kinematics of discrete tracers, as done by \cite{Hartke2020}, which found indeed a lower faction for the unbounded component of diffuse light of $\sim4\%$. Therefore, a lower amount of unbound intra-group diffuse light in the LEO I group could be reliable.
In addition, the average surface brightness profile of NGC~3379 from \citet{Watkins2014} (see left panel of Fig. 6 in \citet{Watkins2014}, for comparison) maps the outermost regions of the galaxy out to $\sim$ 14 arcmin in $B$ band and down to $m_B \sim 29.3$~mag. In this work (see top left panel of Fig. \ref{fig:ellipseNGC3379}), 
we are able to map the surface photometry out to 18.5 arcmin, and down to $m_g \sim 29.4$~mag, which corresponds to $m_B \sim 30.2$~mag \citep{Kostov2018}. Therefore, we are able to integrate the diffuse light contribution over a larger area and $\sim 1$~mag deeper.

\subsection{IGL versus GCs distribution: intra-group baryons}\label{subsec:GC}

\begin{figure*}
\centering
    \resizebox{0.85\textwidth}{0.85\textheight}{\includegraphics{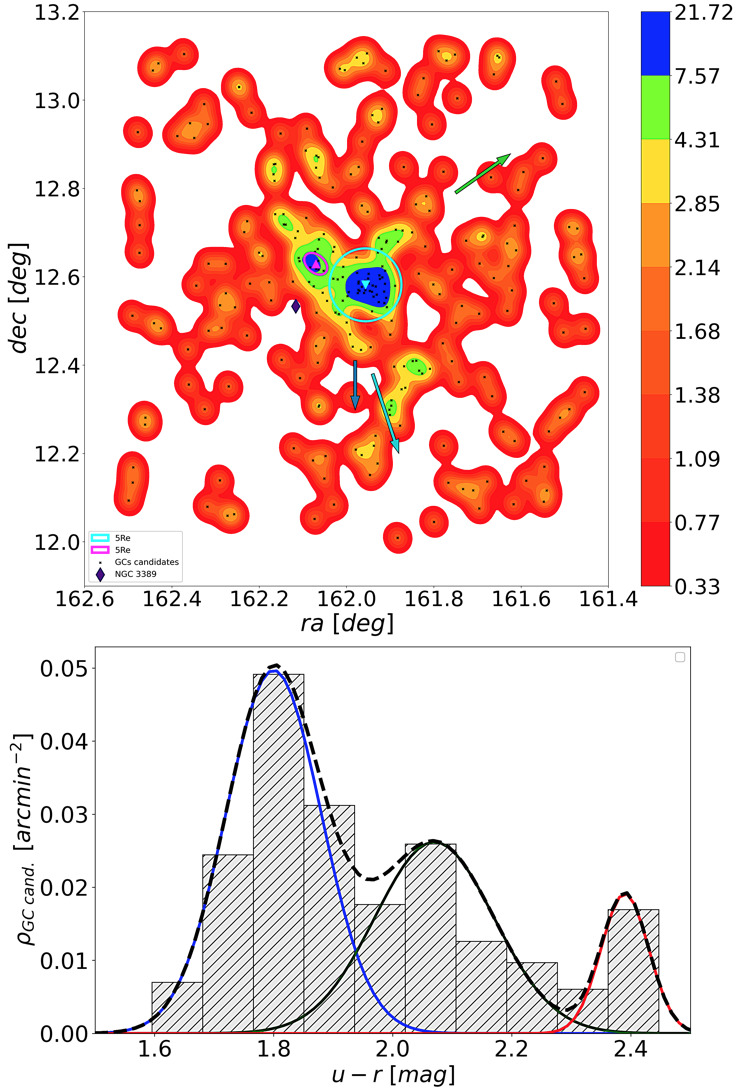}}
    \caption{Top panel: GC candidates 2D density map. The position of NGC\,3379 and NGC\,3384 are marked with cyan and magenta symbols and lines, respectively. A dark-violet diamond shows the position of NGC\,3389.The green arrow shows the direction of the stream of GCs and diffuse light, aligned with the direction of NGC\,3338; the dark-blue arrow shows the direction of the second stream of diffuse light; the ligh-blue arrow marks the direction of M\,96}. Lower  panel: color density distribution (corrected for background contamination, see text) over NGC\,3379. The position of the blue, "green" and red GCs, as derived with GMM, are shown with blue, green and red (solid line) gaussians, respectively, together with the combined fit to the entire population (black dashed line).
    \label{fig:kdeGC}
\end{figure*}

Figure \ref{fig:kdeGC} shows the density distribution of the GCs, using the GCs catalog derived in Sec \ref{subsec:GC analysis}, obtained with a kernel density 
estimator\footnote{We used the KDEpy python package, which implements several kernel density estimators. In particular we used a Gaussian kernel with a $\sigma \sim 0.08$ deg. For the package used in this work see \url{https://seaborn.pydata.org/generated/seaborn.kdeplot.html}}. 
The overdensity of GCs candidates on the LEO\,I pair with respect to the background contamination reveals some surprising features. 
Firstly, we observe that the global GC population, with center on the pair of galaxies, stretches from North-East to South-West with an inclination 
that nearly matches with the line connecting NGC\,3379 and NGC\,3384 to the other bright galaxy in the group, (M\,96) and, in the South-West side, 
overlaps with what seems to be a patch of IGL. Furthermore, nearly orthogonal to such direction, in the side opposite to the irregular NGC\,3389, 
we  find evidence of a GCs over-density stretching along a further possible IGL patch in the same direction.
The presence of these two sub-structures has been verified using different isodensity contours, or independently matched color catalogs 
(e.g. using the $gri$, the $ur$ and the $gi$), and in all cases the presence of both features is confirmed, although with different numbers of GCs
candidates and contaminants. In spite of the small number of GCs expected in the area, and the shallow $u$ band images available, 
the 1:1 correspondence of these features with the IGL patches and with the possible links to interactions with bright companion galaxies in the 
field (M\,96 and NGC\,3338), would actually support them as real features rather than artifacts due to size-of-the-sample effects.
We also inspected the color-distribution of our GCs catalog. Although we expect a large contamination, the presence of relatively 
large portions of sky far from the target galaxies, and with no obvious GCs overdensity related to the LEO\,I pair, allowed us 
to characterize the color distribution of background regions (i.e. of most likely non-GCs sources). Hence, we derived the color 
distribution of the sources on the two galaxies within $\sim 12$ arcmin and, after normalizing to the area, 
subtracted the color-density distributions on and off the galaxies. The top panel of Figure \ref{fig:kdeGC} shows the color density distribution over 
NGC\,3379\footnote{To avoid the contamination of GCs from the bright companion, NGC\,3384, we only considered the area on NGC\,3379 within a circle centered on the photometric barycenter of two galaxies, extending on the half side of NGC\,3379.}. 
The background subtracted color distribution is shown in the botton panel of Figure \ref{fig:kdeGC}. 
The distribution appears clearly multi-modal, with two dominant peaks at $(u{-}r)_{blue}\sim$1.8 mag and $(u{-}r)_{green}$=2.1 mag, with width of $\sigma_{blue}=0.08$ and $\sigma_{green}=0.1$, respectively. The presence of such bi-modal distribution is indeed not unexpected \citep[][]{Larsen2001}. 
We find also a possible third color peak, at $(u{-}r)_{red}\sim2.4$~mag with a $\sigma_{red}=0.04$, 
to whom no reference exist in the literature 
to best of our knowledge. The three peaks are also consistently found in the $g{-}i$. 
Although the latter peak could possibly be a background artefact, by inspecting the positions of the GCs we find a number of red GCs lies close to the galaxy core, where the more metal-rich population of GCs would indeed be expected \citep[][]{KP97,Cantiello2007gc,Harris09}. Moreover, five of the red core-GCs, out of ten, belong to the master catalog.
In summary, although the field of the LEO\,I pair is relatively close and its GCs population has already been targeted by several other studies, 
the availability of the large area high-quality VST images allowed us to identify several GCs system properties previously unknown. 
Further spectroscopic follow-ups would be of great interest to confirm the existence of a direct link between the new features and 
the other galaxies in the region.

\section{Discussion and Conclusions}\label{sec:conc}

In this work we have presented new multi-band ($ugri$) deep imaging data for the LEO I pair of galaxies, as part of the VEGAS project. 
Thanks to the large covered area (3.9 deg$^2$ around the core of the pair) and the long integration time, we map the light distribution 
down to $\mu_g \sim 30$~mag/arcsec$^2$ and out to 63~kpc from the BGG NGC~3379, and we provided a census of the GCs in this system.
The main goal of this work was to derive the amount of IGL and its color, which are fundamental 
indicators of the formation history of the group. 

The main results of this work are summarised below.

\begin{itemize}
    \item The new map of diffuse intra-group light extends out to 63~kpc and presents two very faint ($\mu_g\sim28-29$~mag/arcsec$^2$) streams protruding from NGC~3379 and elongated toward North-West and South.
    \item The fraction of the diffuse light, coming from the stellar envelope plus the IGL, 
    is $\sim 17\%$ in both $g$ and $r$ bands, with an average color of $g-r=0.75\pm0.04$~mag.
    \item  The GCs population stretches both from North-East to South-West and from North-West to South of the pair, in the last case overlapping with the streams of IGL. 
    \item The color distribution of the GCs appears clearly multi-modal, with two dominant peaks at (u-r)=1.8 mag and 2.1 mag, respectively. 
\end{itemize}

\begin{figure}
 \centering
 \resizebox{1\textwidth}{.3\textheight}{\includegraphics{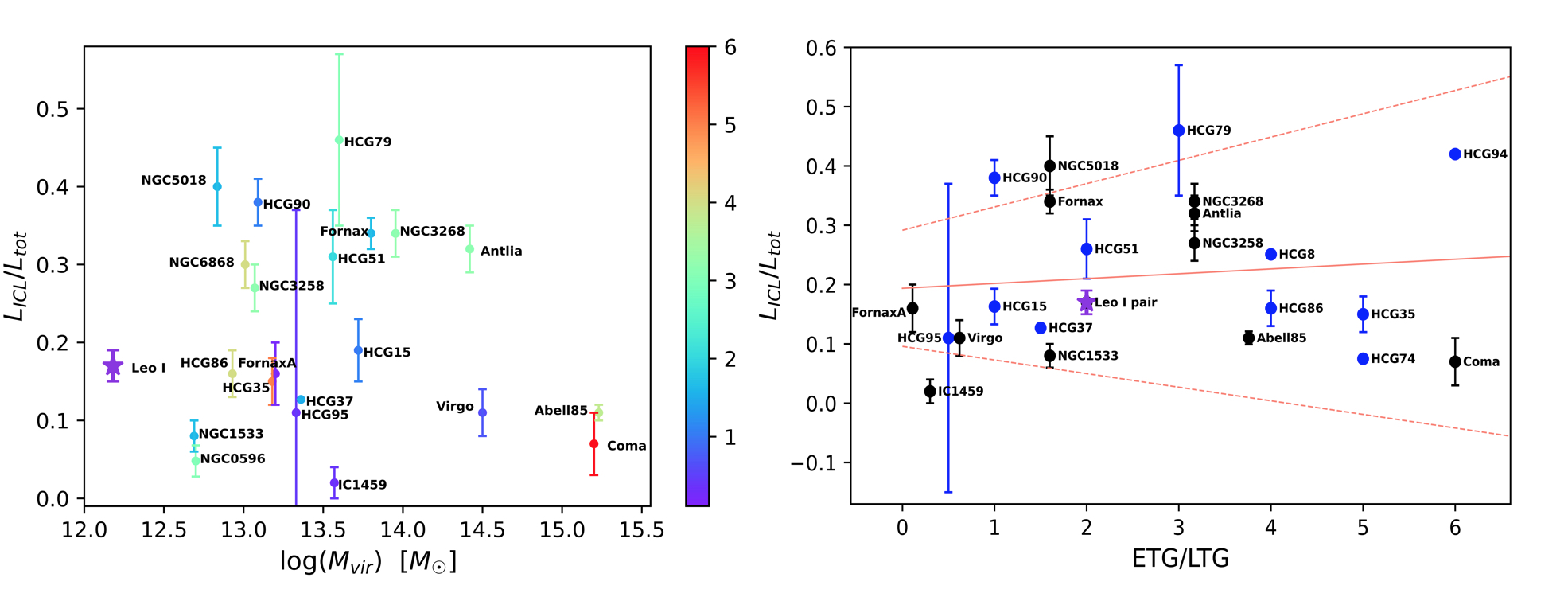}}
    \caption{{\it Left panel:} Luminosity of the diffuse light component normalised to the total luminosity of the environment (group or cluster of galaxies) as a function of the virial mass. 
    {The value derived for LEO I pair is compared with those for several compact groups, taken from \citet{DaRocha2008,Selim2008,Pildis_1995,Poliakov2021}. The estimate for HCG~94 and HCG~37 are provided by \citet{Pildis_1995} and \citet{Poliakov2021} without an error range, respectively.} In addition, values for the IGL estimated for other groups of galaxies derived using VEGAS data are also shown. These are: NGC~5018 group \citep{Spavone2018}, NGC~1533 triplet, IC~1459 group \citep{Iodice_2020}, Fornax A subgroup \citep{Raj_2020}, HCG~86 \citep{RAGUSA2021}, and NGC~6868, NGC~3258,NGC~3268, NGC~0596 (Ragusa et al. 2022, in preparation). The ICL fraction for the Fornax cluster is derived by \citet{Spavone_2020} using FDS data. We also report the values for Virgo \citep{Mihos2017}, Coma \citep{Jim_nez_Teja_2019}, Abell~85 clusters \citep{Brough2017ApJ...844...59B, montes2021buildup} and Antlia cluster (Ragusa et al. 2022, in preparation). Although these estimates are all obtained with the same approach, it must be taken into account that some of them come from different photometric bands. The color of each point is coded according to its ETGs-to-LTGs ratio. It seems to suggest that no evident trend exists between the amount of IGL/ICL and M$_{vir}$. {\it Right panel:} Luminosity of the ICL component normalised to the total cluster or group luminosity as a function of the ETGs-to-LTGs ratio. 
    The IGL for all HCGs available in literature are marked as blue points. Other estimates for groups and clusters of galaxies are indicated with the black points. 
    The value found in this work for LEO I pair is marked with purple star-like point. 
    The coral solid line corresponds to the best fit for the linear correlation and the dashed coral lines mark the 2$\sigma$ significance range of the correlation. 
    The NGC~5018 group and Fornax cluster have an ETG-to-LTG ratio similar to the NGC~1533 group and HCG~37, but former contain about the double amount of IGL-ICL respect to teh latter. The same goes for Coma and HCG~94. In contrast, Virgo and Coma clusters share about the same amount of ICL, but the first has a ETG-to-LTG ratio equal to 0.62, the second 10 times greater  (ETG-to-LTG = 6). The coral solid line corresponds to the best fit for the linear correlation and the dashed coral lines mark the 2$\sigma$ significance range of the correlation.
    The values for HCG~74, HCG~8, HCG~37, and HCG~17 (the latter overlaps to HCG~86) are from \citet[][]{Poliakov2021}. }
    \label{fig:vir_mass}
\end{figure}

The IGL fraction we derived is consistent with estimates already present in other observational
works derived for groups and clusters of galaxies, which ranges between $\sim10\%$ and $45\%$. This is evident
from Fig.~\ref{fig:vir_mass} (left panel), where we plot the fraction of diffuse light as a function 
of the virial mass ($M_{vir}$) of the environment, from several observational studies.
The virial mass and virial radius of LEO I group, centered on NGC~3379, are taken from \citet{Karachentsev2015AstBu..70....1K} 
($M_{vir}$= 1.26 x $10^{13}$~M$_{\odot}$, $R_{vir}$= 467 kpc).
Since our field covers $\sim$ 0.53 $R_{vir}$, we scaled the virial mass according 
to this radius ($M_{vir}$=1.9 x $10^{12}$~M$_{\odot}$), used to derive the $M_{vir}$ for the LEO I pair.
As already pointed out by \citet{RAGUSA2021}, given the
large scatter, Fig.~\ref{fig:vir_mass} further suggests the absence of any trend between the fraction of diffuse light and $M_{vir}$. 
This result is consistent with theoretical predictions from \citet{Sommer-Larsen2006,Monaco_2006,Henriques_2010,Rudick2011,Contini2014}, 
which provide an estimate for the ICL fraction ranging from 20\% to 40\% for any value of $M_{vir}$.
Conversely, an increasing value of the ICL fraction with the virial mass is predicted by several other
theoretical studies, also based on simulations \citep{Purcell2007,Lin_2004,Murante2007,Pillepich2018}.
Based on the distribution of PNe in the halo of NGC~3379, \citet{Hartke2020} estimated a 3.8$\%$ as 
lower limit for the diffuse light in this group. 
Authors found that the PNe profile flattens at about 250~arcsec from the galaxy centre and it is
well approximated by an exponential profile with a scale radius of $r_h = 258 \pm 2$~arcsec.
They suggested that such an excess of PNe can maps an additional diffuse light component in the outskirts of NGC~3379, 
since it differs from the halo of the galaxy for its distinct $\alpha$-parameter value (which is the 
luminosity-specific PNe number) and spatial density distribution.
The transition radius of 250~arcsec, based on the PNe distribution is fully consistent with the transition
radius of 250~arcsec, we derived from the multi-component fit to separate the brightest central part of 
the galaxy from its outskirts (i.e. stellar envelope plus IGL, see Sec.~\ref{subsubsec:fit1d}). 
Furthermore, the scale-length they derived for the exponential profile of the PNe distribution in the
galaxy outskirts is fully consistent with $r_h = 260 \pm 1$~arcsec we obtained from the photometric decomposition. 
The consistency between the photometric and kinematic tracers suggests that the 
estimate of the IGL fraction we derived in this work for the LEO~I pair is robust.
Moreover,  in a recent study, \citet{Hartke2022arXiv220108710H} present a new and extended kinematic survey of PNe around NGC~3379, inside an area of $30'\times 30'$.
They found three distinct kinematic regimes that are linked to different stellar population properties:\
\begin{itemize}
      \item the innermost one, i.e. the "rotating core", extends up to $\sim$ 0.9 arcmin and corresponds to the stellar population formed in situ and metal rich. We identify this innermost component with the first S{\'e}rsic (red curve) presented in the left panel of Fig.~ \ref{fig:ellipseNGC3379}, which also extends up to $\sim$ 1 arcmin;\
      \item the halo, from $\sim$ 0.9 arcmin to $\sim$ 7 arcmin, consisting of a mixture of intermediate-metallicity and metal-rich stars, formed in situ or through major merger events. The scales of this component are consistent with  the second S{\'e}rsic (magenta curve) photometric component we fitted to reproduce the surface brightness profile of NGC~3379 (see Fig.~ \ref{fig:ellipseNGC3379}); \
      \item the exponential envelope, beyond $\sim$ 7 arcmin, composed mostly of metal-poor stars, which traces the IGL component, which turns to be consistent with the exponential stellar envelope found from our three components 
      model of NGC~3379 (see left panel of Fig.~\ref{fig:ellipseNGC3379}), for $R\geq$ 7 arcmin.
\
\end{itemize}

The average color $g-r = 0.75 \pm 0.04$~mag for the IGL in LEO~I pair is comparable with the $g-r$ color estimated
for the diffuse light component in other groups and cluster of galaxies, in the nearby universe 
\citep[$\sim0.7$~mag in the Fornax cluster, $\sim0.8$~mag in HCG~86][]{Iodice2017a,Raj_2020,RAGUSA2021}. 
Moreover, such a value for the IGL color is also consistent with the range of $g-r$ colors predicted for the diffuse light
by \citet[][]{Contini2019}, where $0.7\leq g-r \leq 0.8$~mag at z = 0. In the left panel of the Fig. \ref{fig:vir_mass} we have color-coded the fraction of diffuse light 
in groups and clusters based on their ETGs-to-LTGs ratio. 
In the right panel of the Fig. \ref{fig:vir_mass} we present the fraction of the diffuse light with respect to the total luminosity of the environment as a function of the ETGs-to-LTGs ratio. 
Since the LEO I pair is made up of only one elliptical galaxy and one S0 galaxy,   
$ETGs/LTGs=2$. 
In this plot, the IGL fraction estimated for the LEO~I pair is consistent with the amount of diffuse light expected 
and observed for other environments of comparable ETGs-to-LTGs ratio, as some Hickson Compact groups of galaxies and the 
NGC~1533 triplet.
As already noticed in previous works \citep{DaRocha2008,RAGUSA2021}, a weak trend between the amount of diffuse light and the ETGs-to-LTGs ratio in groups and clusters of galaxies seems to be present. 
Since the more dynamically evolved groups have the largest ETGs-to-LTGs fraction
and the higher probability of tidal interactions, such as stellar stripping and mergers, a 
larger amount of diffuse light is expected for environments dominated by early-type galaxies.
In conclusion, the new deep photometry we have performed and presented in this work revealed an amount of IGL in the 
LEO~I pair which is comparable with the diffuse light in other environments of similar virial mass. 
The estimated photometric transition between the inner and brighter region of the BGG and the diffuse envelope of stellar 
halo plus IGL is consistent with the kinematic one obtained for this target by \citet{Hartke2020} and \citet{Hartke2022arXiv220108710H}. 
The VEGAS deep images have also revealed the presence of two faint stellar streams that might be associated at the IGL 
distribution. The 2-dimensional distribution both of the GCs found in this work and PNe found by \citet{Hartke2020} 
show an over-density overlapping with the two faint stellar streams. This would suggest that they are coherent 
structures of the intra-group baryons. Since they are elongated in the direction of the 
two nearby galaxies M96 and NGC~3338, they could be evidence of gravitational interactions with the pair. 
In particular, the stream in the direction of M96 could result from a head-on collision between LEO group members NGC~3384 
and M96 \citep{Dansac2010ApJ...717L.143M}.





\section*{Acknowledgments}
This work is based on observations collected at the European Southern Observatory (ESO) 
La Silla Paranal Observatory within the VST Guaranteed Time Observations, Programme IDs:  096.B-0582(B), 097.B-0806(A), 090.B-0414(D).
Authors acknowledge financial support from the VST project (P.I. P. Schipani).
ALM acknowledges financial support from the INAF-OAC.
We are grateful to M. Arnaboldi and J. Hartke for the enlightening discussions about the comparison between kinematics and photometry.



\bibliographystyle{frontiersinHLTH&FPHY} 
\bibliography{LEO_group_igl.bib}

    

\begin{table}
\setlength{\tabcolsep}{1.2pt}
\centering
\caption{Properties of the galaxies in the LEO I pair.\\
{\it Note.} Column 1 report the name of the two LEO I pair members. In Col. 2 is given the morphological type. In Cols. 3 and 4 are listed the celestial coordinates of each group member. In Col. 5 is listed the heliocentric radial velocity, and in Col. 6 is reported the distances of the pair members. The distance for NGC~3384 was estimated using the relationship: D = V$_{He}$ H$_0^{-1}$ , with $H_0$ = 73 Km s\textsuperscript{-1} Mpc\textsuperscript{-1} \citep[][]{Riess2018}.} 
\begin{tabular}{lcccccc}
\\
\hline\hline
\\
     Target    &   morphological type      &    R.A.     &  Decl. & Helio. radial velocity & D \\
       &  & J200 & J200 & km s$^{-1}$ & Mpc\\
     (1)&(2)&(3)&(4)&(5)&(6)\\
    \hline
    \\
    $NGC~~3379$ & E1 &10$^{h}$ 47$^{m}$m 49.588 $^{s}$&  +12$^{d}$ 34$^{m}$ 53.85$^{s}$ & 911.07 & 10.23 $\pm$ 0.08 \\
    $NGC~~3384$ & SB0$^{-(s)}$ &10$^{h}$ 48$^{m}$m 16.886 $^{s}$&  +12$^{d}$ 37$^{m}$ 45.38$^{s}$ & 703.9 & 9.64 \\

    \hline
   
    \end{tabular}
    \label{tab:parameterNED}
   
\end{table}

\begin{table}
\setlength{\tabcolsep}{1pt}
\centering
\caption{Structural parameter obtained for the galaxies in the LEO I pair from the fit of the isophotes.\\
{\it Note.} Column 1 report the name of the two pair members. In Col. 2 is reported the total magnitude in the $g$ band. In Cols. 3 and 4 are listed the magnitude at the transition radius in $g$ and $r$ bands, while in Cols. 5 and 6 are given the effective radius in the $g$ band, in arcsec and kpc respectively. The average $g-r$ color is listed in Col. 7. Magnitudes and colors were corrected for Galactic extinction using the extinction coefficients provided by \citet{Schlafly_2011}.} 
\begin{tabular}{lccccccc}
\\
\hline\hline
\\
     Galaxy     &  $m_g$   & $m_g [R\leq R_{tr}]$ &~~$m_r [R\leq R_{tr}]$ &  $R_{e,g}$ &~~ $R_{e,g}$  &  g-r \\
     & [mag] &[mag] & [mag]& [arcsec] &[kpc]& [mag]\\
     (1)&(2)&(3)&(4)&(5)&(6)&(7)\\
    \hline
    \\
    $NGC~~3379$ & ~~9.28 $\pm$ 0.04 &~~ 9.31 $\pm$ 0.01 &~~ 8.51 $\pm$ 0.01 & 86 & 4.3 &~~ 0.82 $\pm$ 0.06\\
    $NGC~~3384$ &~~10.42 $\pm$ 0.01 &~~ &~~& 22 & 1.1 &~~ 0.76 $\pm$ 0.02\\

    \hline
   
    \end{tabular}
    \label{tab:parameterisofote}
   
\end{table}

\begin{table}
\setlength{\tabcolsep}{0.5pt}
\centering
\caption{Structural parameters derived from the 1D fit of the azimuthally averaged surface brightness profiles of the LEO I pair.\\
{\it Note.} Columns 2, 3, 4, 5, 6, 7 report the effective surface brightness, effective
radius and S{\'e}rsic index for the two inner component of each fit, in the g band, whereas columns 8 and 9 list the
central surface brightness and scale length for the outer exponential component. Columns 9 and 10 give a transition radii, in arcsec and kpc scale respectively, derived from the 1D fit of the azimuthally averaged surface brightness profiles of NGC~3379 (see Sec. \ref{subsubsec:fit1d})} 
\begin{tabular}{lccccccccccc}
\\
\hline\hline
\\
  Object&$\mu_{e,1}^g$& $R_{e,1}^g$ & $n_{1}^g$ &$\mu_{e,2}^g$  & $R_{e,2}^g$ & $n_{2}^g$ & $\mu_{0,g}$ & $r_{h,g}$ & $R_{tr,g}$ &$R_{tr,g}$ \\ 
    & [mag/arcsec$^{2}$]&[arcsec]& &[mag/arcsec$^{2}$] & [arcsec]& &[mag/arcsec$^{2}$] & [arcsec]&[arcsec]& [kpc]\\
  (1) & (2) &(3)& (4) & (5) & (6)& (7)& (8)& (9)& (10)& (11)\\
    \hline
    \\
    $NGC3379$ & 18.00$\pm$0.02& 5.1$\pm$ 0.1 & 1.16 &21.45$\pm$ 0.03 & 55.0 $\pm$ 0.5 &2.10 & 24.95$\pm$ 0.02& 260$\pm$1 &252& 12.6\\
    $NGC3384$ & 17.85 $\pm$ 0.01 & 6.0$\pm$ 0.3 & 1.42 &21.75$\pm$ 0.03 & 70 $\pm$ 1 &0.65 & & && \\

    \hline
   
    \end{tabular}
    \label{tab:parameterfit}
   
\end{table}

\begin{table}[h] 

\centering        
\begin{tabular}{l c c c}    
\hline\hline    
\\ 
 Parameters & \ Min &  \ Max  \\ [0.5ex]    
\hline                  
\\
\ \ \ \raisebox{1.5ex}{FWHM} & \raisebox{1.5ex}{0.73} & \raisebox{1.5ex}{0.95}

\\
\ \ \ \ \ \ \raisebox{1.5ex}{C.I.} & \raisebox{1.5ex}{0.45} & \raisebox{1.5ex}{0.57} 
 
\\
\ \raisebox{1.5ex}{Elongation} & \raisebox{1.5ex}{0.92} & \raisebox{1.5ex}{1.28}

\\
\  \raisebox{1.5ex}{Flux radius} & \raisebox{1.5ex}{0.4} & \raisebox{1.5ex}{0.75}

\\
\ \ \ \raisebox{1.5ex}{$\Delta$ (g-i)} &  & \raisebox{1.5ex}{0.3}

\\
 \ \ \ \raisebox{1.5ex}{$\Delta$ (u-r)} &  & \raisebox{1.5ex}{0.5}
 \\
 \ \ \ \raisebox{1.5ex}{$m_{r}$} & \raisebox{1.5ex}{19} & \raisebox{1.5ex}{26}

\\[1ex] 
 
\hline    
\hline   
\end{tabular} 
\caption{Photometric and morphometric parameters adopted for source selections. }

\label{tab:selGC} 
\end{table} 
 
\newpage
\begin{sidewaystable}[htpb]

\centering
\begin{adjustbox}{width=\textwidth}
\begin{tabular}{l c c c c c c c c c c c c c c c c }    
\hline
\hline  
\\ 
R.A.  &Dec.  &$m_r$ &$\Delta m_r$ &$m_g$ & $\Delta m_g$ &$m_u$ &$\Delta m_u$ &$m_i$ &$\Delta m_i$ &Elon. &$Flux_{R}$ &$FWHM$ &$C_S $&C.I. \\
deg (J2000)& deg (J2000)& mag&mag&mag&mag&mag&mag&mag&mag&pixel&pixel& & mag\\
(1)&(2)&(3)&(4)&(5)&(6)&(7)&(8)&(9)&(10)&(11)&(12)&(13)&(14)&(15)\\
\\
\hline
\\
161.8827105 &12.0089343 &20.807 &0.008 &21.449 &0.012 &22.756 &0.102 &20.455 &0.031 &1.206 &2.441 &3.79 &0.979 &0.47 \\
161.7994979 &12.0450087 &22.259 &0.024 &22.914 &0.034 &24.417 &0.442 &21.906 &0.075 &1.038 &2.366 &4.15 &0.941 &0.477 \\
162.272812 &12.0591517 &21.201 &0.009 &21.857 &0.013 &22.852 &0.094 &20.934 &0.03 &1.034 &2.669 &3.76 &0.927 &0.481\\
161.6174692 &12.0853106 &21.631 &0.012 &22.205 &0.015 &23.569 &0.171 &21.294 &0.033 &1.099 &2.532 &3.69 &0.952 &0.48\\
161.5480706 &12.0908925 &19.197 &0.002 &19.774 &0.002 &20.887 &0.016 &18.913 &0.005 &1.074 &2.839 &3.92 &0.859 &0.502\\
161.5412847 &12.1161331 &20.89 &0.006 &21.499 &0.008 &22.845 &0.078 &20.563 &0.017 &1.061 &2.657 &4.05 &0.906 &0.508\\
161.7331405 &12.1196303 &19.755 &0.002 &20.251 &0.002 &21.381 &0.022 &19.437 &0.008 &1.04 &2.799 &3.86 &0.868 &0.492\\
161.7668362 &12.1338537 &20.501 &0.004 &21.001 &0.005 &22.237 &0.048 &20.117 &0.013 &1.068 &2.856 &4.12 &0.744 &0.517\\
162.4898132 &12.1341857 &21.529 &0.014 &22.141 &0.018 &23.315 &0.155 &21.293 &0.04 &1.146 &2.878 &4.18 &0.925 &0.502\\
161.6778305 &12.1365518 &21.673 &0.01 &22.173 &0.013 &23.36 &0.116 &21.273 &0.032 &1.081 &2.67 &3.66 &0.879 &0.474\\
&$.$&$.$&$.$&$.$&$.$&$.$&$.$&$.$&$.$&$.$&$.$&$.$&$.$&$.$&$.$\\
&$.$&$.$&$.$&$.$&$.$&$.$&$.$&$.$&$.$&$.$&$.$&$.$&$.$&$.$&$.$\\
&$.$&$.$&$.$&$.$&$.$&$.$&$.$&$.$&$.$&$.$&$.$&$.$&$.$&$.$&$.$\\
&$.$&$.$&$.$&$.$&$.$&$.$&$.$&$.$&$.$&$.$&$.$&$.$&$.$&$.$&$.$\\

\hline   
\hline   
\end{tabular}
\end{adjustbox}
\caption{Selected GCs catalog : Col. (1-2) R.A and Dec. in degrees (J2000), Col. (3-4) magnitudes and uncertainties in the $r$ band, Col. (5-6) magnitudes and uncertainties in the $g$ band, Col. (7-8) magnitudes and uncertainties in the $u$ band, Col. (9-10) magnitudes and uncertainties in the $i$ band, Col. (11) is the Elongation parameter from SExtractor defined as the ratio between the major and minor axis of the sources, Col. (12) is the Flux Radius parameter from SExtractor which is the radius containing half of the flux from the sources, Col. (13) is the FWHM parameter from SExtractor, Col. (14) is the Class Star  parameter from SExtractor which classify sources, Col. (15) is the Concentration Index defined as the difference between magnitude at difference aperture, in particular we did the difference between 4-pixel aperture magnitude and 6-pixel one. }
\label{tab:GC} 
\end{sidewaystable}

 \begin{sidewaystable*}

 \centering

 \fontsize{11}{9}\selectfont
 \begin{tabular}{l c c c c c c c c c c c c c c c c c c }    
 \hline\hline    
\\

 R.A.       &Dec.  &$m_r$ &$\Delta m_r$ & $m_g$ &$\Delta m_g$ &$m_u$ &$\Delta m_u$ &$m_i$ &$\Delta m_i$ &Elon. &F.R. &FWHM &C.S.&C.I & T$_p$ &T$_s$ &T$_x$ \\
 deg (J2000)& deg (J2000)& mag& mag       & mag   & mag       & mag   & mag   & mag   & mag     &  & pixel &    pixel&       & mag  &   &   &  \\
 (1)    & (2)   & (3) &(4)& (5)&(6)&(7)&(8)&(9)&(10)&(11)&(12)&(13)&(14)&(15)&(16)&(17)&(18)\\
 \hline
 \\

 161.903091 &12.5899022 &21.249 &0.006 &21.874 &0.008 &22.893 &0.059 &20.926 &0.025 &1.121 &3.507 &5.4 &0.029 &0.629 &1 &0 &0\\
 161.98628 &12.5574202 &20.234 &0.003 &20.811 &0.004 &22.007 &0.027 &19.887 &0.014 &1.262 &2.726 &5.28 &0.931 &0.569 &1 &2 &0\\
 162.055516 &12.7069587 &21.218 &0.005 &21.881 &0.007 &23.368 &0.091 &20.841 &0.025 &1.116 &3.618 &4.82 &0.033 &0.579 &0 &2 &0\\
 161.973382 &12.5818206 &20.884 &0.007 &21.466 &0.008 &22.675 &0.048 &20.594 &0.024 &1.162 &3.092 &4.75 &0.323 &0.556 &1 &2 &0\\
 161.94939 &12.5773175 &21.133 &0.012 &21.773 &0.013 &23.348 &0.092 &20.823 &0.023 &1.032 &2.822 &4.7 &0.969 &0.515 &1 &0 &0\\
 162.002186 &12.5464099 &20.745 &0.004 &21.374 &0.005 &22.673 &0.049 &20.417 &0.018 &1.089 &3.327 &4.68 &0.137 &0.542 &1 &2 &0\\
 161.913558 &12.5406199 &19.786 &0.002 &20.439 &0.002 &21.787 &0.023 &19.438 &0.008 &1.18 &3.281 &4.57 &0.219 &0.534 &1 &2 &0\\
 162.057819 &12.6206602 &21.124 &0.008 &21.981 &0.012 &23.745 &0.128 &20.668 &0.022 &1.116 &2.783 &4.57 &0.921 &0.567 &0 &2 &0\\
 162.065532 &12.5984584 &21.477 &0.007 &22.282 &0.011 &24.112 &0.178 &20.971 &0.028 &1.102 &2.845 &4.47 &0.135 &0.554 &0 &2 &0\\
 161.918018 &12.6234968 &21.518 &0.007 &22.136 &0.01 &23.375 &0.102 &21.264 &0.033 &1.122 &3.17 &4.43 &0.138 &0.541 &1 &2 &0\\
 &$.$&$.$&$.$&$.$&$.$&$.$&$.$&$.$&$.$&$.$&$.$&$.$&$.$&$.$&$.$&$.$&$.$&$.$\\
 &$.$&$.$&$.$&$.$&$.$&$.$&$.$&$.$&$.$&$.$&$.$&$.$&$.$&$.$&$.$&$.$&$.$&$.$\\
 &$.$&$.$&$.$&$.$&$.$&$.$&$.$&$.$&$.$&$.$&$.$&$.$&$.$&$.$&$.$&$.$&$.$&$.$\\
 &$.$&$.$&$.$&$.$&$.$&$.$&$.$&$.$&$.$&$.$&$.$&$.$&$.$&$.$&$.$&$.$&$.$&$.$\\

 \hline    
 \hline
 \end{tabular}
 \caption{Master catalog parameter: Col. (1-2) R.A and Dec. in degrees (J2000), Col. (3-4) magnitudes and uncertainties in the $r$ band,Col. (5-6) magnitudes and uncertainties in the $g$ band, Col. (7-8) magnitudes and uncertainties in the $u$, Col. (9-10) magnitudes and uncertainties in the $i$ band, Col. (11) is the Elongation parameter from SExtractor defined as the ratio between the major and minor axis of the sources, Col. (12) is the Flux Radius parameter from SExtractor which is the radius containing half of the flux from the sources, Col. (13) is the FWHM parameter from SExtractor, Col. (14) is the Class Star  parameter from SExtractor which classify sources, Col. (15) is the Concentration Index defined as the difference between magnitude at difference aperture, in particular we did the difference between 4-pixel aperture magnitude and 6-pixel one. Col. (16-18) refer to literature classifications of globular cluster, $T_p$ stand for photometric,$T_s$ stands for spettroscopy, $T_x$ stand for LMXB studies. The full table is available at the VEGAS project web pages, and in the CDS repository.}
 \label{tab:masterGC} 
 \end{sidewaystable*}

\end{document}